\documentclass[twoside,12pt]{article}
\usepackage{epsfig}

\newcommand{\be}{\begin{equation}}
\newcommand{\ee}{\end{equation}}
\newcommand{\bea}{\begin{eqnarray}}
\newcommand{\eea}{\end{eqnarray}}

\begin{document}

\title{Equation of state for $\beta$-stable hot nuclear matter}

\author{Ch.C. Moustakidis and C.P. Panos\\
$^{}$ Department of Theoretical Physics, Aristotle University of
Thessaloniki, \\ 54124 Thessaloniki, Greece }

\maketitle

\begin{abstract}
We provide an equation of state for hot nuclear matter in
$\beta$-equilibrium by applying a momentum-dependent effective
interaction. We focus on the study of the equation of state of
high-density and high-temperature nuclear matter, containing
leptons (electrons and muons) under the chemical equilibrium
condition in which neutrinos have left the system. The conditions
of charge neutrality and equilibrium under $\beta$-decay process
lead first to the evaluation of proton and lepton fractions and
afterwards  of internal energy, free energy, pressure and in total
to the equation of state of hot nuclear matter. Thermal effects on
the properties and equation of state of nuclear matter are assesed
and analyzed in the framework of the proposed effective
interaction model. Special attention is dedicated to the study of
the contribution of the components of $\beta$-stable nuclear
matter to the entropy per particle, a quantity of great interest
for the study of structure and collapse of
supernova.\vspace{0.3cm}

PACS number(s): 21.65.+f, 21.30.Fe,
24.10.Pa, 26.60.+c, 26.50.+x \\

Keywards: Hot Nuclear Matter, Effective Interaction, Equation of
State, Nuclear Symmetry Energy, Proton Fraction, Neutron Star.
\end{abstract}

\section{Introduction}
The equation of state (EOS) of hot nuclear matter determines the
structure inside a  supernova \cite{Bethe-90} and a hot neutron
star \cite{Prakash-97,Li-06,Li-08} and affects the state of
matter, such as its chemical composition. In addition, the
equation of state plays important roles for the study of the
supernova explosion, as well as on determining the evolution of a
neutron star at the birth stage. The profiles of a neutron star as
the density, temperature and proton fraction during the cooling,
which affect the reaction rate of neutrino process inside the
star, are determined through the equation of state.
%Hence, there is a strong need to provide  a
%reliable EOS for the supernova explosion and the birth of neutron
%stars.

There exist many calculations for hot nuclear matter with
applications to the properties of hot neutron stars and supernova
\cite{Bethe-90,Prakash-97,Li-06,Li-08,Baym-71,Bethe-79,Friedman-81,Lattimer-91,
Takatsuka-94,Takatsuka-96,Das-07,Haar-86,Bombaci-95,Kupper-74,Lattimer-78,Eid-80,Lamb-79,Antia-80,Onsi-94,
Sumiyoshi-94,Lattimer-85,Kanzawa-07,Modarres-97,Manka-00,Zuo-03,Chen-01,
Mishra-93,Ccernai-92,Lee-01,Muller-95,Baldo-99,Burgio-07,Xu-07-3,Wang-00,Jena-04,Samaddar-96,
Tolos-07,Alla-93,Moustakidis-07,Moustakidis-08}. G. Baym {\it et
al.} provided an EOS of neutron matter \cite{Baym-71} and Bethe
{\it et al.} \cite{Bethe-79} an EOS for the gravitational collapse
of stars. Friedman and Pandharipande \cite{Friedman-81} performed
variational calculations of the equation of state of hot and cold,
nuclear and neutron matter. Lattimer and Swesty carried out
calculations of the EOS for stellar collapse, using the
compressible liquid-drop model for nuclei \cite{Lattimer-91}. M.
Prakash {\it et al.} \cite{Prakash-97} investigated the structure
of neutron stars shortly after their birth, by applying various
nuclear models. Takatsuka {\it et al.}
\cite{Takatsuka-94,Takatsuka-96} have performed detailed
calculations for supernova matter, within the framework of finite
temperature Hartree-Fock approach, with effective nucleon-nucleon
interaction. Recently Das {\it et al.} \cite{Das-07} have
calculated the EOS of dense supernova matter within the finite
temperature Brueckner Goldstone approach with effective two-body
Sussex interaction.

%The last years there is an increasing interest for the study of
%hot nuclear matter properties with application for supernova
%evaluation and hot neutron stars structure and cooling \cite{}.

The present work is based on the previous work of Prakash {\it et
al.} \cite{Prakash-97}. More specifically, in order to study the
properties and the EOS of hot nuclear matter, we apply a
momentum-dependent effective interaction model (MDIM), which is
able to reproduce the results of more microscopic calculations of
dense matter at zero temperature and which can be extended to
finite temperature
\cite{Prakash-97,Moustakidis-07,Moustakidis-08}.

The aim of this work is to apply a momentum-dependent interaction
model for the study of  hot nuclear matter EOS under
$\beta$-equilibrium. The present model has the additional
property, compared to the previous ones, that the temperature
affects not only the kinetic part of the energy density, but  also
influences  the interaction part of the energy density as well. In
that way, we are able to study simultaneously thermal effects not
only on the kinetic part of the symmetry energy and symmetry free
energy, but also on the interaction part of the above quantities.
This is important in the sense  that the density dependent
behavior of the symmetry energy and symmetry free energy influence
strongly the values of the proton fraction and as a consequence
the composition of  hot $\beta$-stable nuclear matter, under
consideration.

Using the above method , we will show that the thermal energy (and
also the related quantities) depend sensitively on the momentum
dependence of the nuclear interaction. We concentrate our study on
the properties of hot nuclear matter in the density range
$n_0<n<6n_0$ (where $n_0=0.16$ fm$^{-3}$ is the saturation
density) and temperature range $0<T<30$ MeV, taking into account
that nuclear matter consists of neutrons, protons, electrons and
muons with their relative concentrations determined from the
conditions of charge neutrality and equilibrium under
$\beta$-decay process in the absence of neutrino trapping.

The article is organized as follows. In Sec. II the model and
relative formulas are discussed and analyzed. Results are reported
and discussed in Sec. III, whereas the summary of the work is
given in Sec. IV.

%%%%%%%%%%%%%%%%%%%%%%%%%
\section{The model}
%%%%%%%%%%%%%%%%%%%%%%%%%%%
We start by outlining the momentum dependent interaction model,
then we define the thermodynamic quantities of nuclear matter and
finally we analyze the  $\beta$-equilibrium conditions, the
contribution on pressure and energy of leptons and the total
equation of state of nuclear matter.
%%%%%%%%%%%%%%%%%%%%%%%%%%%%%%%%%%%%%%%%%%%
\subsection{Momentum dependent interaction model }
%%%%%%%%%%%%%%%%%%%%%%%%%%%%%%%%%%%%%%%%%%%%%%%%%
The schematic potential model, employed here, is designed to
reproduce the results of the microscopic calculations of both
nuclear and neutron-rich matter at zero temperature and can be
extended to finite temperature \cite{Prakash-97}. The energy
density of the asymmetric nuclear matter (ANM) is given by the
relation
\begin{equation}
\epsilon(n_n,n_p,T)=\epsilon_{kin}^{n}(n_n,T)+\epsilon_{kin}^{p}(n_p,T)+
V_{int}(n_n,n_p,T), \label{E-D-1}
\end{equation}
where $n_n$ ($n_p$) is the neutron (proton) density and the total
baryon density is $n=n_n+n_p$. The contributions of the kinetic
parts are
\begin{equation}
\epsilon_{kin}^n(n_n,T)+\epsilon_{kin}^p(n_p,T)=2 \int \frac{d^3
k}{(2 \pi)^3}\frac{\hbar^2 k^2}{2m}
\left(f_n(n_n,k,T)+f_p(n_p,k,T) \right), \label{E-K-D-1}
\end{equation}
where $f_{\tau}$, (for $\tau=n,p$) is the Fermi-Dirac distribution
function with the form
\begin{equation}
f_{\tau}(n_{\tau},k,T)=\left[1+\exp\left(\frac{e_{\tau}(n_{\tau},k,T)-\mu_{\tau}(n_{\tau},T)}{T}\right)
\right]^{-1}. \label{FD-1}
\end{equation}
The nucleon density $n_{\tau}$ is evaluated from the following
integral
\begin{equation}
n_{\tau}=2 \int \frac{d^3k}{(2\pi)^3}f_{\tau}(n_{\tau},k,T)=2 \int
\frac{d^3k}{(2\pi)^3}\left[1+\exp\left(\frac{e_{\tau}(n_{\tau},k,T)-\mu_{\tau}(n_{\tau},T)}{T}\right)\right]^{-1}.
\label{D-1}
\end{equation}
In Eq. (\ref{FD-1}), $e_{\tau}(n_{\tau},k,T)$ is the single
particle energy (SPE) and $\mu_{\tau}(n_{\tau},T)$ stands for the
chemical potential of each species. The SPE has the form
\begin{equation}
e_{\tau}(n_{\tau},k,T)=\frac{\hbar^2k^2}{2m}+U_{\tau}(n_{\tau},k,T),
\label{esp-1}
\end{equation}
where the single particle potential $U_{\tau}(n_{\tau},k,T)$ is
obtained by the functional derivative  of the interaction part of
the energy density with respect to the distribution function
$f_{\tau}$. Including the effect of finite-range forces between
nucleons, to avoid acausal behavior at high densities, the
potential contribution is parameterized as follows
\cite{Prakash-97}
\begin{eqnarray}
V_{int}(n_n,n_p,T)&=&\frac{1}{3}An_0\left[\frac{3}{2}-(\frac{1}{2}+x_0)I^2\right]u^2
+\frac{\frac{2}{3}Bn_0\left[\frac{3}{2}-(\frac{1}{2}+x_3)I^2\right]u^{\sigma+1}}
{1+\frac{2}{3}B'\left[\frac{3}{2}-(\frac{1}{2}+x_3)I^2\right]u^{\sigma-1}}
\nonumber \\ &+& u \sum_{i=1,2}\left[C_i \left({\cal J}_n+{\cal
J}_p\right)+\frac{(C_i-8Z_i)}{5}I\left({\cal J}_n-{\cal
J}_p\right)\right], \label{V-all}
\end{eqnarray}
where
\begin{equation}
{\cal J}_{\tau}= \ 2 \int \frac{d^3k}{(2\pi)^3}
g(k,\Lambda_i)f_{\tau}(n_{\tau},k,T). \label{J-tau}
\end{equation}
%%%%%%%%%%%%%%%%%%%%%%%%%%%%%%%%%%%%%%%%%%%%%%%%%%%%%%%%%%%%%%%%%%%%%%%%%%%%%%%%%%%%%%%%%%%%%%%%%%%%%%%%%%%%%%%%%%%
%%%%%%%%%%%%%%%%%%%%%%%%%%%%%%%%%%%%%%%%%%%%%%%%%%%%%%%%%%%%%%%%%%%%%%%%%%%%%%%%%%%%%%%%%%%%%%%%%%%%%%%%%%%%%%%%%%%

In Eq.~(\ref{V-all}), $I$ is the asymmetry parameter
($I=(n_n-n_p)/n$) and $u=n/n_0$, with $n_0$ denoting the
equilibrium symmetric nuclear matter density, $n_0=0.16$
fm$^{-3}$. The asymmetry parameter $I$ is related to the proton
fraction $Y_p$ by the equation $I=(1-2Y_p)$. The parameters $A$,
$B$, $\sigma$, $C_1$, $C_2$ and $B'$, which appear in the
description of symmetric nuclear matter, are determined in order
that $E(n=n_0)-mc^2=-16$ {\rm MeV}, $n_0=0.16$ fm$^{-3}$, and the
incompressibility to be $K=240$ {\rm MeV}.  The additional
parameters $x_0$, $x_3$, $Z_1$, and $Z_2$, which are used to
determine the properties of asymmetric nuclear matter, are treated
as parameters constrained by empirical knowledge
\cite{Prakash-97}. The parameterizations used in the present model
have only a modest microscopic foundation. Nonetheless, they have
the merit of being able to closely approximate more physically
motivated calculations as presented in Fig.~1. More precisely, in
Fig.~1 we compare the energy per baryon (for symmetric nuclear
matter (Fig.~1a) and pure neutron matter (Fig.~1b)) calculated by
the present schematic model (MDIM), with those of existent, state
of the art calculations by Wiringa et al. \cite{Wiringa-88-b} and
Pandharipande et al. \cite{Pandharipande-98}.

The first two terms of the right-hand side of Eq.~(\ref{V-all})
arise from local contact nuclear interaction which lead to power
density contributions as in the standard Skyrme equation of state.
The first one concerns attractive interaction while the second one
is repulsive, and both are assumed to be temperature independent.
The third term describes the effects of finite range interactions
according to the chosen function $g(k,\Lambda_i)$, and is the
temperature dependent part of the interaction. This interaction is
attractive and important at low momentum, but it weakens and
disappears at very high momentum.  The function $g(k,\Lambda_i)$,
suitably chosen to simulate finite range effects, has the
following form \cite{Prakash-97}
\begin{equation}
 g(k,\Lambda)=\left[1+\left(\frac{k}{\Lambda_{i}}\right)^2
\right]^{-1}, \label{g-1} \end{equation}
where the finite range parameters are $\Lambda_1=1.5 k_F^{0}$ and
$\Lambda_2=3 k_F^{0}$ and $k_F^0$ is the Fermi momentum at the
saturation point $n_0$.

The main origin of the momentum dependence in Brueckner theory is
the nonlocality of the exchange interaction. Following the
discussion of Bertsch et al. \cite{Bertsch-88} a single-particle
potential $U(n)$ which depends only on the baryon density is
oversimplified. What is more, it is well known that nuclear
interaction has strong exchange effects which give rise to a
momentum dependence in the single-particle potential and as a
consequence has an effect on the energy density functional. The
question here is how best to parameterize the momentum dependence
in modelling the potential $U(n,k)$. A promising approach might be
to adopt the relativistic mean field model, where
$U(n,k)=U_{\nu}n+\frac{U_sn}{\sqrt{1+k^2/m^2}}$. The above
potential exhibits a strong momentum dependence for small $k$
which diminishes to zero at high momentum. In order to perform
extensive studies in heavy ion collision studies, Gale et al.
\cite{Gale-87}  have proposed  the following parametrization for
the momentum part of the single-particle
\[U(n,k)\sim C \frac{n}{n_0}\frac{1}{1+({\bf k}-\langle{\bf
k}'\rangle)^2/\Lambda^2}.
\]
This has a proper fall-off at high $k$ and Galilean invariance is
assured by measuring $k$ with respect to the average of the
particles in the neighborhood, $\langle{\bf k}'\rangle$. For
static nuclear matter we have $\langle{\bf k}'\rangle=0$.

The present model, which is a generalization of that proposed by
Gale et al. \cite{Gale-87}, has been successfully applied
 in heavy ion collisions and astrophysical studies over the years
\cite{Prakash-97,Li-06,Li-08,Mishra-93,Ccernai-92,Xu-07-3,Prakash-88,BaoLi-03}.

In order to clarify the relative contribution of the three terms
of the potential energy density mentioned above, we plot them as a
function of the baryon density in Fig.~2a. In this figure we have
that
\begin{eqnarray}
V_A&=&\frac{1}{3}An_0\left[\frac{3}{2}-(\frac{1}{2}+x_0)I^2\right]u^2,                 \nonumber\\
V_B&=&\frac{\frac{2}{3}Bn_0\left[\frac{3}{2}-(\frac{1}{2}+x_3)I^2\right]u^{\sigma+1}}
{1+\frac{2}{3}B'\left[\frac{3}{2}-(\frac{1}{2}+x_3)I^2\right]u^{\sigma-1}},                      \\
V_C&=& u \sum_{i=1,2}\left[C_i \left({\cal J}_n^i+{\cal
J}_p^i\right)  + \frac{(C_i-8Z_i)}{5}I\left({\cal J}^i_n-{\cal
J}_p^i\right)\right].                  \nonumber \label{3-terms}
\end{eqnarray}
As mentioned above, the first term $V_A$ corresponds to an
attractive interaction, whereas the second one $V_B$ corresponds
to repulsive interaction and dominates for high values of $n$
($n>0.6$ fm$^{-3}$). Both of these terms are temperature
independent. The third term $V_C$ (is plotted for $T=0$) contains
the momentum dependent part of the interaction, corresponds to
attractive interaction, and its main contribution is to compete
with the repulsive interaction of $V_B$ for high values of $n$ and
as a consequence avoid acausal behavior of the EOS  at high
densities. The term $V_C$ consists of two finite range terms, one
corresponding to a long-range attraction and the other to a
short-range repulsion.

Thermal effects on the momentum dependent term $V_C$ are displayed
in Fig.~2b. The contribution of $V_C$ is plotted for various
values of $T$. It is therefore concluded that thermal effects are
more pronounced for high values of $T$ ($T>10$ MeV), leading to a
less attractive contribution. More precisely, we find that for
small values of $n$ (i.e. $n=0.15$ fm$^{-3}$ ) $V_C$ increases
(compared to the cold case $T=0$) $3\%-20\%$ for $T=10-30$. For
higher values of $n$ the increase is even less.

An additional test for the present model is to compare the single
particle potential $U_{\tau}(n_{\tau},k,T)$ (or
$U_{\tau}(n,I,k,T)$) originated from the present version of the
momentum dependent interaction with other calculations. The single
particle potential $U_{\tau}(n,I,k,T)$ (protons or neutrons),
obtained from the functional derivative of the interaction part of
the energy density (Eq.~(\ref{V-all})) with respect to the
distribution function $f_{\tau}$,  has the general form
\cite{Moustakidis-08}
\begin{equation}
U_{\tau}(n,I,k,T)=U_{\tau}^A(n,I)+U_{\tau}^B(n,I)+U_{\tau}^{MD}(n,I,k,T).
\label{U-total}
\end{equation}
It is of interest to see that the single particle potentials are
separated into two parts. The first one,
$U_{\tau}^A(n,I)+U_{\tau}^B(n,I)$ is a function only of the baryon
density $n$ and the isospin asymmetry parameter $I$. The second
one, $U_{\tau}^{MD}(n,I,k,T)$ has an additional dependence on $T$
and $k$. Actually, $U_{\tau}^{MD}(n,I,k,T)$ is mainly responsible
for the trend of the effective mass and also the effective mass
splitting. Additionally, it is connected with the effect of the
temperature on the interacting  part of the energy density
\cite{Moustakidis-08}.

The single-particle potential in symmetric nuclear matter has been
calculated microscopically for several Hamiltonians by Wiringa
\cite{Wiringa-88-2}. These Hamiltonians include nucleon-nucleon
potentials fit to scattering data  and three nucleon potentials
fit to binding energies of few-body nuclei and saturation
properties of nuclear matter.  The potential was parameterized
using the ansatz
\begin{equation}
U(n,k)=\alpha(n)+\frac{\beta(n)}{1+(\frac{k}{\Lambda(n)})^2},
\label{wiri-1}
\end{equation}
where the density dependent parameters $\alpha(n)$, $\beta(n)$ and
$\Lambda(n)$, for three types  Hamiltonians,  are listed in Table
I of Ref.~\cite{Wiringa-88-2}.

Furthermore, the single-particle potential has been derived by Li
et al. \cite{Li-93}. The derivation is based upon the Bonn
meson-exchange model for the nucleon-nucleon interaction and the
Dirac-Brueckner approach for nuclear matter. The potential, named
DBHF, has been parameterized as following
\begin{equation}
U(n,k)=\alpha n+\beta n^{\gamma}+\delta\ln^2(\epsilon (\hbar c
k)^2+1)n^{\sigma}. \label{Li-pot}
\end{equation}
The parameters $\alpha$, $\beta$, $\gamma$, $\delta$, $\epsilon$
and $\sigma$ are listed  in Table I of Ref.~\cite{Li-93}.

A comparison with the results of UV14+TNI, UV14+UVII, AV14+UVII
and DBHF interactions show that (see Fig.~3) $U(n,k)$ (for $T=0$)
obtained from the present model is very reasonable for at least up
to the value $k=3$ fm$^{-1}$. The agreement is not so good for
high values of $k$, especially compared with the DBHF interaction,
but as has been pointed out by Li et al.~\cite{Li-93}, the
parametrization in Eq.~\ref{Li-pot} is bad for large $k$ since it
continues to grow with increasing $k$, while the exact potential
becomes independent of $k$ for large momenta. In conclusion, the
present results show that the momentum dependent interaction
model, which has been applied in the present work,  provides a
reliable representation of $U(n,k)$ for a wide range of density
and momentum.

%%%%%%%%%%%%%%%%%%%%%%%%%%%%%%%%%%%%%%%%%%%%%%%%%%%%%%%%%%%%%%%%%%%%%%%%%%%%%%%%%%%%%%%%%%%%%%%%%%%%%
%%%%%%%%%%%%%%%%%%%%%%%%%%%%%%%%%%%%%%%%%%%%%%%%%%%%%%%%%%%%%%%%%%%%%%%%%%%%%%%%%%%%%%%%%%%%%%%%%%%%%

The energy density of asymmetric nuclear matter at density $n$ and
temperature $T$, in a good approximation, is expressed as
\begin{equation}
\epsilon(n,T,I)=\epsilon(n,T,I=0)+\epsilon_{sym}(n,T,I),
\label{e-asm-1}
\end{equation}
where
\begin{equation}
\epsilon_{sym}(n,T,I)=nI^2 E_{sym}^{tot}(n,T)=n I^2
\left(E_{sym}^{kin}(n,T)+E_{sym}^{int}(n,T)\right).
\label{e-sym-1}
\end{equation}
In Eq.~(\ref{e-sym-1}) the nuclear symmetry energy
$E_{sym}^{tot}(n,T)$ is separated in   two parts corresponding to
the kinetic contribution $E_{sym}^{kin}(n,T)$ and the interaction
one $E_{sym}^{int}(n,T)$.

From Eqs.~(\ref{e-asm-1}) and (\ref{e-sym-1}) and setting $I=1$,
we find that the nuclear symmetry energy $E_{sym}^{tot}(n,T)$ is
given by
\begin{equation}
E_{sym}^{tot}(n,T)=\frac{1}{n}\left(\epsilon(n,T,I=1)-\epsilon(n,T,I=0)
\right). \label{Esym-d-1}
\end{equation}
Thus, from Eq.~(\ref{Esym-d-1}) and by a suitable choice of the
parameters $x_0$, $x_3$, $Z_1$ and $Z_2$, we can obtain various
forms for the density dependence of the symmetry energy
$E_{sym}^{tot}(n,T)$.

It is well known that the need to explore different forms for
$E_{sym}^{tot}(n,T)$ stems from the uncertain behavior at high
density \cite{Prakash-97}.   The high-density behavior of symmetry
energy is the least known property of dense matter
\cite{Kutschera-94,Li-02,Fuchs-06}, with different nuclear models
giving contradictory predictions. Thus, in relativistic mean field
(RMF) models, the symmetry energy increases  strongly with the
density of nuclear matter \cite{Glendenning-97}, while in many
realistic potential models of nuclear matter in the variational
approach \cite{Friedman-81,Wiringa-88}, the symmetry energy
saturates and then bends over at higher densities.

Recently, the density dependence of the symmetry energy in the
equation of state of isospin asymmetric nuclear matter has been
studied using isoscaling of the fragment yields and the
antisymmetrized molecular dynamic calculation \cite{Shetty-07}. It
was observed that the experimental data at low densities are
consistent with the form of symmetry energy, $E_{sym}(u)\approx
31.6u^{0.69}$, in close agreement with those predicted by the
results of variational many-body calculations. In
Ref.~\cite{Shetty-07} it was suggested also that the heavy ion
studies favor a dependence of the form $E_{sym}(u)\approx
31.6u^{\gamma}$, where $\gamma=0.6-1.05$. This constrains the form
of the density dependence of the symmetry energy at higher
densities, ruling out an extremely "stiff" and "soft" dependence
\cite{Shetty-07}.

Additionally, Chen et al.~\cite{Chen-05} also showed, using the
isospin dependent Boltzmann-Uehling-Uhlenbeck transport model
calculations, that a stiff density dependence of the symmetry
energy parameterized as, $E_{sym}(u)\approx 31.6u^{1.05}$ explains
well the isospin diffusion data \cite{Tsang-04} from NSCL-MSU
(National Superconducting Cyclotron Laboratory at Michigan State
University).

In this paper, since we are interested mainly in the study of
thermal effects on the nuclear symmetry energy, we choose a
specific form for it, enabling us to reproduce accurately the
results of many other theoretical studies
\cite{Lee-98,Sammarruca-08}. In Ref.~\cite{Lee-98} the authors
carried out a systematic analysis of the nuclear symmetry energy
in the formalism of the relativistic Dirac-Brueckner-Hartree-Fock
approach, using the Bonn one-boson-exchange potential. In a very
recent work \cite{Sammarruca-08}, the authors applied a similar
method as in Ref.~\cite{Lee-98} for the microscopic predictions of
the equation of state in asymmetric nuclear matter. In that case
$E_{sym}(u)$ is obtained employing the simple parametrization
$E_{sym}(u)=C u^{\gamma}$ with $\gamma=0.7-1.0$ and $C\approx 32$
MeV.  The authors conclude that a value of $\gamma$ close to $0.8$
gives a reasonable description of their predictions, although the
use of different functions in different density regions may be
best for an optimal fit \cite{Sammarruca-08}. The results of
Refs.~\cite{Lee-98,Sammarruca-08}  are well reproduced by
parameterizing the nuclear symmetry energy according to the
formula
\begin{equation}
E_{sym}^{tot}(n,T=0)= \underbrace{13
u^{2/3}}_{Kinetic}+\underbrace{17
F(u)}_{Interaction}.\label{Esym-3}
\end{equation}
For the function $F(u)$, which parametrizes the interaction part
of the symmetry energy, we apply the following form
\begin{equation}
F(u)=u. \label{Fu-form}
\end{equation}
The parameters $x_0$, $x_3$, $Z_1$ and $Z_2$ are chosen so that
Eq.~(\ref{Esym-d-1}), for $T=0$, reproduces the results of
Eq.~(\ref{Esym-3}) for  the function $F(u)=u$.

In one of our previous paper~\cite{Psonis-07}, the potential part
of the symmetry energy has been parameterized in the generalized
form, $F(u)=u^c$, and the obtained nuclear equations of state are
applied to the systematic study of the global properties of a
neutron star (masses, radii and composition). We obtained a linear
relation between the parameter $c$ and the radius and the maximum
mass of the neutron star~\cite{Psonis-07}. Additionally, we found
that a linear relation between the radius and the derivative of
the symmetry energy near the saturation density $n_0$ holds
 \cite{Psonis-07}.

It is worthwhile to point out that the above parametrization of
the interacting part of the nuclear symmetry energy is used
extensively for the study of neutron star properties
\cite{Prakash-97,Prakash-94}, as well as for the study of the
collisions of neutron-rich heavy ions at intermediate energies
\cite{Li-97,Baran-05}. For a very recent review of the
applications of the proposed momentum dependent effective
interaction model and its specific parameterizations, see
Ref.~\cite{Li-08} (and references therein).

%%%%%%%%%%%%%%%%%%%%%%%%%%%%%%%%%%%%%%%%%%%%%%%%%%%%%%%
\subsection{Thermodynamic description of hot nuclear matter}
%%%%%%%%%%%%%%%%%%%%%%%%%%%%%%%%%%%%%%%%%%%%%%%%%%%%%%%%%%%%
In order to study the properties of nuclear matter at finite
temperature, we need to introduce the Helmholtz free energy $F$.
The differential of the total free energy $F_{tot}$ (the total
free energy of baryons contained in volume $V$) and total internal
energy $E_{tot}$ (the total internal energy of baryons contained
in volume $V$) are given by \cite{Goodstein-85,Fetter-03}
\begin{equation}
{\rm d} F_{tot}=-S_{tot}{\rm d}T-P{\rm d}V+\sum_i \mu_i {\rm d}N_i
\label{d-F}
\end{equation}
\begin{equation}
{\rm d} E_{tot}=T{\rm d}S_{tot}-P{\rm d}V+\sum_i \mu_i {\rm d}N_i
\label{d-E}
\end{equation}
where $S_{tot}$ is the total entropy of the baryons, while $\mu_i$
and $N_i$ are the chemical potential and the number of particles
of each species respectively.

It is easy to prove that the free energy per particle $F$ is
written as \cite{Goodstein-85,Fetter-03}
\begin{equation}
F(n,T,I)=E(n,T,I)-TS(n,I,T). \label{Free-1}
\end{equation}
In Eq.~(\ref{Free-1}),  $E$ is the internal energy per particle,
$E=\epsilon/n$, and $S$ is the  entropy per particle, $S=s/n$.
From Eq.~(\ref{Free-1}) is also concluded that for $T=0$, the free
energy $F$ and the internal energy $E$ coincide.

The entropy density $s$ has the same functional form as that of a
non interacting gas system, given by the equation
\begin{equation}
s_{\tau}(n,I,T)=-2\int \frac{d^3k}{(2\pi)^3}\left[f_{\tau} \ln
f_{\tau}+(1-f_{\tau}) \ln(1-f_{\tau})\right].  \label{s-den-1}
\end{equation}
The total internal energy $E_{tot}$  is useful for studying
isentropic processes. In that description of a thermodynamic
system, the pressure and the chemical potential are defined as
follows \cite{Goodstein-85,Fetter-03}
\begin{equation}
P=-\left(\frac{\partial E_{tot}}{\partial
V}\right)_{S,N_i}=n^2\left(\frac{\partial \epsilon/n}{\partial
n}\right)_{S,N_i}, \qquad \qquad \qquad
\mu_{i}=\left(\frac{\partial E_{tot}}{\partial
N_i}\right)_{S,V,N_{j\neq i}}=\left(\frac{\partial
\epsilon}{\partial n_i}\right)_{S,V,n_{j\neq i}}. \label{P-m-E}
\end{equation}

Now we are going to study the properties and the EOS of nuclear
matter by considering an isothermal process. In that, the pressure
and the chemical potential are connected with the derivative of
the total free energy $F_{tot}$. More specifically, they are
defined as follows
\begin{equation}
P=-\left(\frac{\partial F_{tot}}{\partial V}
\right)_{T,N_i}=n^2\left(\frac{\partial f/n}{\partial
n}\right)_{T,N_i}, \qquad \qquad \mu_{i}=\left(\frac{\partial
F_{tot}}{\partial N_i}\right)_{T,V,N_{j\neq
i}}=\left(\frac{\partial f}{\partial n_i}\right)_{T,V,n_{j\neq
i}}, \qquad \qquad \label{P-m-F}
\end{equation}
where $f$ is the free energy density.

The pressure $P$ can be  calculated also from equations
\cite{Goodstein-85,Fetter-03}
\begin{equation}
VP=TS_{tot}-E_{tot}+\sum_{i}\mu_iN_i,\qquad {\rm or} \qquad \qquad
P=Ts-\epsilon+\sum_{i}\mu_in_i. \label{P-1}
\end{equation}

It is also possible to calculate the entropy per particle $S(n,T)$
by differentiating the free energy density $f$ with respect to the
temperature
\begin{equation}
S(n,T)=- \left(\frac{\partial f/n}{\partial T} \right)_{V,N_i}.
\label{S-dif-f}
\end{equation}

The comparison of the two entropies, that is from
Eqs.~(\ref{s-den-1}) and (\ref{S-dif-f}), provides a test of the
approximation used in the present work.

It is easy to show by applying Eq.~(\ref{P-m-F}) that (see  for a
proof \cite{Prakash-94} as well as \cite{Burgio-07})
\begin{eqnarray}
\mu_n&=&F+u\left(\frac{\partial F}{\partial
u}\right)_{Y_p,T}-Y_p\left(\frac{\partial F}{\partial
Y_p}\right)_{n,T}, \nonumber
\\ \mu_p&=&\mu_n+\left(\frac{\partial F}{\partial
Y_p}\right)_{n,T}, \nonumber \\
\hat{\mu}&=&\mu_n-\mu_p=-\left(\frac{\partial F}{\partial
Y_p}\right)_{n,T}.
 \label{mu-p-n}
\end{eqnarray}
%\begin{equation}
%\hat{\mu}=\mu_n-\mu_p=-\left(\frac{\partial F}{\partial Y_p}
%\right)_{n,T}, \label{mhat1}
%\end{equation}
%
We can define the symmetry free energy per particle $F_{sym}(n,T)$
by the following parabolic approximation (see also
\cite{Burgio-07,Xu-07-3})
\begin{equation}
F(n,T,I)=F(n,T,I=0)+I^2F_{sym}(n,T)=
F(n,T,I=0)+(1-2Y_p)^2F_{sym}(n,T),\label{Free-Parabolic}
\end{equation}
where
\begin{equation} F_{sym}(n,T)= F(n,T,I=1)-F(n,T,I=0).
\label{Free-asym}
\end{equation}
It is worthwhile to notice that the above approximation is not
valid from the beginning, but one needs to check the validity of
the parabolic law in the present model before using it. As we see
later, that law is well satisfied as well as the parabolic law
holding for the energy.

Now, by applying Eq.~(\ref{Free-Parabolic}) in Eq.~(\ref{mu-p-n}),
we obtain the  key relation
\begin{equation}
\hat{\mu}=\mu_n-\mu_p=4(1-2Y_p)F_{sym}(n,T). \label{mhat-2}
\end{equation}
The above equation is similar to that obtained for cold nuclear
matter by replacing $E_{sym}(n)$ with $F_{sym}(n,T)$.
%%%%%%%%%%%%%%%%%%%%%%%%%%%%%%%%%%%%%%%%%%%%%%
\subsection{$\beta$-equilibrium, leptons contribution and equation of state}

%In $\beta$-stable matter the processes
Stable high density nuclear matter must be in chemical equilibrium
for all types of reactions, including the weak interactions, while
$\beta$ decay and electron capture take place simultaneously
\begin{equation}
n \longrightarrow p+e^{-}+\bar{\nu}_e, \qquad \qquad p +e^{-}
\longrightarrow n+ \nu_e.
\end{equation}

Both types of reactions change the electron per nucleon fraction,
$Y_e$ and thus affect the equation of state. Here, we assume that
neutrinos generated in those reactions have left the system. The
absence of neutrino-trapping has a dramatic effect on the equation
of state and mainly induces a significant change on the values of
the proton fraction $Y_p$ \cite{Takatsuka-94,Takatsuka-96}. The
absence of neutrinos implies that
\begin{equation}
\hat{\mu}=\mu_n-\mu_p=\mu_e. \label{chem-1}
\end{equation}

When the energy of electrons is large enough (i.e. greater than
the muon mass), it is energetically favorable for the electrons to
convert to muons
\begin{equation}
e^{-} \longrightarrow \mu^{-}+\bar{\nu}_{\mu}+\nu_e .
\end{equation}
Denoting the muon chemical potential by $\mu_{\mu}$, the chemical
equilibrium established by the above process and its inverse is
given by
\[\mu_e=\mu_{\mu}.\]
Taking into account that the threshold for muons occurs for
$\mu_{\mu}=m_{\mu}c^2\simeq 105.7$ MeV, one may expect muons to
appear roughly at nuclear density $n=0.16$ fm$^{-3}$.

Thus, in total, we consider that nuclear matter contains neutrons,
protons, electrons, and muons. They are in a $\beta$-equilibrium,
where the following relations hold
\begin{equation}
\mu_n=\mu_p+\mu_e, \qquad \qquad \mu_e=\mu_{\mu}. \label{b-1}
\end{equation}

Furthermore, they obey the charge neutrality condition i.e.
\begin{equation}
n_p=n_e+n_{\mu}. \label{charge}
\end{equation}

The leptons (electrons and muons) density is given by the
expression
\begin{equation}
n_l=\frac{2}{(2\pi)^3}\int \frac{ {\rm d}{\bf
k}}{1+\exp\left[\frac{\sqrt{\hbar^2k^2c^2+m_l^2c^4}-\mu_l}{T}
\right]}.
%=\frac{1}{\pi^2}\int_0^{\infty} \frac{ k^2 {\rm
%dk}}{1+\exp\left[\frac{\sqrt{\hbar^2k^2c^2+m_l^2c^4}-\mu_l}{T}
%\right]}.
\label{n-lepton-1}
\end{equation}
\\
One can solve self-consistently
Eqs.~(\ref{mhat-2}),(\ref{b-1}),(\ref{charge}) and
(\ref{n-lepton-1}) in order to calculate the proton fraction
$Y_p$, the lepton fractions $Y_e$ and $Y_{\mu}$, as well as the
electron chemical potential $\mu_e$ as a function of the baryon
density $n$, for various values of the temperature $T$.

The next step is to calculate the energy and pressure of leptons
given by the following formulae
\begin{equation}
\epsilon_{l}(n_l,T)=\frac{2}{(2\pi)^3}\int \frac{\sqrt{\hbar^2 k^2
c^2+m_l^2c^4} \ {\rm d}{\bf
k}}{1+\exp\left[\frac{\sqrt{\hbar^2k^2c^2+m_l^2c^4}-\mu_l}{T}
\right]},
%=\frac{1}{\pi^2}\int_0^{\infty} \frac{\sqrt{\hbar^2 k^2
%c^2+m_l^2c^4}\ k^2 {\rm
%d}k}{1+\exp\left[\frac{\sqrt{\hbar^2k^2c^2+m_l^2c^4}-\mu_l}{T}
%\right]},
\label{e-lepton-1}
\end{equation}
%%%%%%%%%%
\begin{equation}
P_l(n_l,T)=\frac{1}{3}\frac{2(\hbar c)^2}{(2\pi)^3}\int
\frac{1}{\sqrt{\hbar^2 k^2 c^2+m_l^2c^4}}\frac{\ k^2 \ {\rm d}{\bf
k}}{1+\exp\left[\frac{\sqrt{\hbar^2k^2c^2+m_l^2c^4}-\mu_l}{T}
\right]}.
% \nonumber \\
%&=&\frac{(\hbar c)^2}{3\pi^2}\int_0^{\infty}
%\frac{1}{\sqrt{\hbar^2 k^2 c^2+m_l^2c^4}}\frac{ k^4 \ {\rm d}
%k}{1+\exp\left[\frac{\sqrt{\hbar^2k^2c^2+m_l^2c^4}-\mu_l}{T}
%\right]}.
\label{P-lepton-1}
\end{equation}
The  chemical potentials of electrons and muons are equal and
according to Eqs.~(\ref{mhat-2}) and (\ref{b-1}) are
\begin{equation}
\mu_{e}=\mu_{\mu}=\mu_p-\mu_n=4\left(1-2Y_p(n,T)\right)F_{sym}(n,T)=4I(n,T)F_{sym}(n,T).
\label{chem-lep-1}
\end{equation}
%
%So, if ones know the function $Y_p(n,T)$ (or $I(n,T)$) and
%$E_{sym}^{tot}(n,T)$ can calculate lepton fraction ($Y_{e}$,
%$Y_{\mu}$), energy density and pressure of the lepton as a
%function of the baryon density, for various values of $T$.

The equation of state of hot nuclear matter in $\beta$-equilibrium
(considering that it consists of neutrons, protons, electrons and
muons) can be obtained by calculating the total energy density
$\epsilon_{tot}$ as well as the total pressure $P_{tot}$. The
total energy density is given by
\begin{equation}
\epsilon_{tot}(n,T,I)=\epsilon_b(n,T,I)+\sum_{l=e,\mu}\epsilon_l(n,T,I),
\label{e-de-1}
\end{equation}
where $\epsilon_b(n,T,I)$ and $\epsilon_l(n,T,I)$ are the
contributions of baryons and leptons respectively. The total
pressure is
\begin{equation}
P_{tot}(n,T,I)=P_b(n,T,I)+\sum_{l=e,\mu}P_l(n,T,I), \label{Pr-1}
\end{equation}
where $P_b(n,T,I)$ is the contribution of the baryons (see
Eq.~(\ref{P-1})) i.e.
\begin{equation}
P_b(n,T,I)=T\sum_{\tau=p,n}s_{\tau}(n,T,I)+\sum_{\tau=n,p}n_{\tau}\mu_{\tau}(n,T,I)-\epsilon_b(n,T,I),
\label{Pr-2}
\end{equation}
while $P_l(n,T,I)$ is the contribution of the leptons (see
Eq.~(\ref{P-lepton-1})). From Eqs.~(\ref{e-de-1}) and (\ref{Pr-1})
we can construct the isothermal curves for energy and pressure and
finally derive the isothermal behavior of the equation of state of
hot nuclear matter under $\beta$-equilibrium.

\section{Results and Discussion}
The schematic potential model, which has been applied in the
present work, has been designed to reproduce the results of the
more microscopic calculations of both nuclear and neutron-rich
matter up to high values of baryon density (see Fig.~1). The
behavior of the high density EOS is of great significance to the
determination of hot protoneutron stars and cold neutron stars
structure. The model has the additional advantage that with the
appropriate parametrization, is able to reproduce different forms
of the density dependence of the nuclear symmetry energy.

In view of the above discussion we calculate the equation of state
of hot asymmetric nuclear matter by applying a momentum dependent
effective interaction model describing the baryons interaction. We
consider that nuclear matter contains neutrons, protons, electrons
and muons under $\beta$-equilibrium and charge neutrality. The key
quantities in our calculations are the proton fraction $Y_p$ and
also the asymmetry free energy defined in Eq.~(\ref{Free-asym}).

In order to check the validity of the parabolic approximation
(\ref{Free-Parabolic}), we plot in Fig.~4 the difference
$F(n,T,I=1)-F(n,T,I=0)$ as a function of $(1-2Y_p)^2$ at
temperature $T=10$ and $T=30$ MeV for three baryon densities,
i.e., $n=0.2$ fm$^{-3}$, $n=0.3$ fm$^{-3}$, and $n=0.4$ fm$^{-3}$.
It is seen that in a good approximation an almost linear relation
holds between $F(n,T,I=1)-F(n,T,I=0)$ and $(1-2Y_p)^2$. A similar
behavior of $F_{sym}(n,T)$ is found by Xu {\it et al.}
\cite{Xu-07-3}, applying an isospin and momentum dependent
interaction model.

It is worth to present the calculation recipe of our work. The
outline of our approach is the following: For a fixed baryon
density $n$, temperature $T$, and asymmetry parameter $I$,
Eq.~(\ref{D-1}) may be solved iteratively in order to calculate
the quantity
\begin{equation}
\eta_{\tau}(n_{\tau},T)=\frac{\mu_{\tau}(n_{\tau},T)-\tilde{U}_{\tau}(n_{\tau},T)}{T},
\label{eta-1}
\end{equation}
where
\begin{equation}
\tilde{U}_{\tau}(n_{\tau},T)=U_{\tau}(n_{\tau},k,T)-\tilde{U}_{\tau}(n_{\tau},k).
\label{U-tau}
\end{equation}
%(both for protons and neutrons).

Knowledge of $\eta_{\tau}(n,T)$ allows the evaluation of
$\tilde{U}_{\tau}(n_{\tau},T)$, which then may  be employed to
infer the chemical potential from
\begin{equation}
\mu_{\tau}(n_{\tau},T)=T\eta_{\tau}(n_{\tau},T)+\tilde{U}(n_{\tau},T),
\label{Chem-1}
\end{equation}
required as an input for the calculation of the Fermi-Dirac
distribution function  $f_{\tau}(n_{\tau},k,T)$. The knowledge of
$f_{\tau}(n_{\tau},k,T)$ permits the calculation of the bulk
quantities of asymmetric nuclear matter.

$F_{sym}(n,T)$, for various values of the temperature $T$, was
derived with a least-squares fit to the numerical values according
to Eq.~(\ref{Free-asym}) and has the form
\begin{eqnarray}
F_{sym}(u;T=0)&=& 13 u^{2/3}+17u \nonumber\\
F_{sym}(u;T=5)&=&3.653+28.018u-1.512u^2+0.185u^3-0.001u^4,
\nonumber \\
F_{sym}(u;T=10)&=&5.995+26.157u-0.827u^2+0.068u^3-0.002u^4, \nonumber \\
F_{sym}(u;T=20)&=&13.200+21.267u+0.800u^2-0.193u^3+0.014u^4,
\nonumber \\
F_{sym}(u;T=30)&=&21.087+17.626u+1.645u^2-0.289u^3+0.018u^4.
 \label{Fsym-T-fit}
\end{eqnarray}
where the case with $T=0$, is included as well. In that case
$F_{sym}$ coincides with $E_{sym}$.

In Fig.~5 we present the behavior of the free energy,
corresponding to hot $\beta$-stable nuclear matter, as a function
of the baryon density $n$, for various values of the temperature
$T$. It is obvious that the thermal effects are more pronounced
for low values of the density $n$.

In Fig.~6 we plot the calculated free energy for symmetric nuclear
matter and pure neutron matter of the proposed momentum dependent
interaction model in comparison with the values of the free energy
calculated by Friedman and Pandharipande (FP model)
\cite{Friedman-81}. In the FP model the equation of state of hot
and cold nuclear and neutron matter has been calculated in the
framework of a variational calculation, where a realistic nuclear
interaction containing two- and three-nucleon body nucleon-nucleon
interaction has been used. In the case of symmetric nuclear matter
the results of the two models are very similar up to values
$n=0.4-0.5$ fm$^{-3}$ depending on the values of $T$. The above
agreement is expected, in the sense that a part of the parameters
of the  model applied in the present work are determined from
constraints provided by the empirical properties of symmetric
nuclear matter at the equilibrium density $n_0=0.16$ fm$^{-3}$.

However, there is an obvious disagreement in the case of pure
nuclear matter, where in the  two models the free energy exhibits
a different trend, especially for higher values of $n$. The above
disagreement will be explained below.

In Fig.~7 we display the internal energy per particle
$E(n,T)=\epsilon(n,T)/n$ given by Eq.~(\ref{E-D-1}) for various
values of temperature. Thermal effects, as expected, are more
pronounced for low values of the baryon density $n$ and less
important for high values of $n$.

In Fig.~8 we display the internal energy of  symmetric nuclear
matter and pure neutron matter, for $T=0$, calculated by the MDIM
and FP models. In accordance with the case of the free energy,
there is a very good agreement in symmetric nuclear matter, but an
obvious disagreement is exhibited in pure neutron matter. The
explanation of the  agreement in the first case is the same as in
the case of the free energy. The disagreement is due to  the
completely different behavior of the two models of the nuclear
symmetry energy, presented in Fig.~9. In our model the parameters
$x_0$, $x_3$, $Z_1$ and $Z_2$  chosen so that
Eq.~(\ref{Esym-d-1}), for $T=0$, reproduce the results of
Eq.~(\ref{Esym-3}) for  the function $F(u)=u$. Consequently,
$E_{sym}(n)$ shows an increasing trend shown in Fig.~9. In
contrast, in the FP model, $E_{sym}(n)$ is a slightly increasing
function of $n$ for low $n$ and then  a decreasing function of $n$
for $n>0.5$ fm$^{-3}$.

In addition, we plot the nuclear symmetry energy  extracted from
experimental results and presented in Ref.~\cite{Shetty-07}, where
$E_{sym}(u)$ is parameterized according to the relation
$E_{sym}(u)\approx 31.6u^{0.69}$ as well as experimental results
extracted from Ref.~\cite{Chen-05}, where $E_{sym}(u)$ is given by
$E_{sym}(u)\approx 31.6u^{1.05}$. The important point to be noted
is that both cases clearly favor a stiff density dependence of the
symmetry energy at higher densities, ruling out the very stiff and
very soft predictions. These results can thus be employed to
constrain the form of the density dependence of the symmetry
energy at supranormal densities relevant for neutron star studies
\cite{Shetty-07}. In the same figure  the theoretical predictions
of Ref.~\cite{Sammarruca-08} are presented, where $E_{sym}(u)$ is
parameterized by $E_{sym}(u)\approx 32u^{0.8}$.

The results of Ref.~\cite{Shetty-07} are in a good agrement with
the present model up to $n=0.3$ fm$^{-3}$ while the theoretical
predictions of Ref.~\cite{Sammarruca-08} are  very close to the
present model up to very high values of the baryon density $n$.

However, our motivation, here, is not to perform a systematic
comparison of various models, but we would like just to present
the similarities and the deviations existing between the models.
The deviations, concerning the symmetry energy behavior of the two
models (MDIM and FP model) are well reflected on the behavior of
the free energy and internal energy of pure neutron matter as
shown in Figs.~6 and 8.

In Fig.~10 we plot the thermal energy per particle
\[E_{thermal}(n,T)=E(n,T)-E(n,T=0),\] of $\beta$-stable matter as a
function of the baryon density $n$ for various values of
temperature $T$. The most striking feature of $E_{thermal}(n,T)$
is that for small values of $T$, the thermal contribution to the
internal energy is almost independent of the density $n$. For high
values of $T$ the situation is different and $E_{thermal}(n,T)$,
for fixed values of $T$, is a decreasing function of the density
$n$.

$E_{thermal}(n,T)$ can be decomposed to separate contributions of
the kinetic and potential energies as follows:
\[E_{thermal}(n,T)=E_{thermal}^{kin}(n,T)+E_{thermal}^{pot}(n,T).\]
We find that for small values of the baryon density (i.e. $n=0.2$
fm$^{-3}$) the contribution, to $E_{thermal}(n,T)$ of the
potential energy $E_{thermal}^{pot}(n,T)$ is $20 \%-10\% $ for
$T=5-30$ MeV. For medium values of $n$ (i.e. $n=0.4$ fm$^{-3}$) is
$43 \%-20\% $ for $T=5-30$ MeV and for higher values (i.e. $n=0.6$
fm$^{-3}$) is $70 \%-30\% $ for $T=5-30$ MeV. Hence, it is
concluded that the potential part of the energy (as a result of
the momentum dependence of the interaction) contributes
significantly to the thermal energy, mainly for small values of
$T$ (for fixed values of $n$) and for large values of $n$ (for
fixed values of $T$).

At this point, it is worthwhile to compare the results for the
pressure obtained by applying Eqs.~(\ref{P-1}) and (\ref{P-m-F}).
Thus, in Fig.~11 we plot $P$ of asymmetric nuclear matter for
$Y_p=0.1$ and $0.3$ at $T=10$ and $30$ MeV. The full lines give
the results calculated from Eq.~(\ref{P-1}), while the squares
give results obtained by differentiating $F(n,T)$
(Eq.~(\ref{P-m-F})). The two calculations for the pressure are in
excellent agreement. This agreement provides a test of the
calculations performed in the present model.

It is of interest also to study the effect of the temperature on
the baryon pressure defined by equation (\ref{P-1}). A related
quantity is the thermal pressure $P_{thermal}(n,T)$  defined as:
\[P_{thermal}(n,T)=P(n,T)-P(n,T=0).\]
$P_{thermal}(n,T)$ as a function of $n$, for various values of $T$
is seen in Fig.~12. $P_{thermal}(n,T)$, in all of the cases, is an
increasing function of the baryon density.

The proton fraction affects the reaction rate of neutrino process
inside that star. If a neutron star has a large proton fraction,
the cooling rate may drastically change through the high neutrino
emissivity due to the direct Urca process. This process can occur
if the proton fraction in the matter of a cold neutron star
exceeds the critical value of $0.11$-$0.15$ and would lead to the
rapid cooling of the neutron star. Thus, it is important to
calculate the proton fraction as a function of the baryon density
and investigate the temperature effects on that.

Fig.~13 displays the fractions of protons, electrons and muons as
functions of the density, for various values of $T$. The proton
fraction is an increasing function of $T$ and this effect is more
pronounced for $T>10$ MeV. The proton fraction $Y_p$ was derived
also with a least-squares fit to the numerical results obtained
from our calculations, leading to the following relations (for
$n>0.15$ fm$^{-3}$).
\begin{eqnarray}
Y_p(n;T=0)&=&-0.050+0.633n-0.521n^2+0.184n^3, \nonumber\\
Y_p(n;T=5)&=&-0.046+0.625n-0.514n^2+0.179n^3, \nonumber \\
Y_p(n;T=10)&=&-0.032+0.570n-0.436n^2+0.139n^3, \\
Y_p(n;T=20)&=&0.021+0.378n-0.163n^2+0.004n^3, \nonumber\\
Y_p(n;T=30)&=&0.109-0.062n+0.908n^2-1.270n^3+0.580n^4. \nonumber
\label{Y-ptoton-T}
\end{eqnarray}

In Fig.~14 we plot the Fermi distribution function $f_{p,n}(n,T)$
both for neutrons and protons for various values of $T$. We
observe that the diffuseness of $f_p(n,T)$ is larger than that of
$f_n(n,T)$. We give an explanation (see also \cite{Takatsuka-94}):
the ratio of $T$ to the Fermi kinetic energy $\epsilon_{Fi}$  is a
measure of the thermal effect. Thus by comparing the two ratios we
have (see also Appendix)
\[\left(\frac{(T/\epsilon_{Fp})}{(T/\epsilon_{Fn})}
\right)={k_F^n}^2/{k_F^p}^2=Y_n^{3/2}/Y_{p}^{3/2}. \] But, due to
$Y_n>Y_p$, we conclude that we expect  the diffuseness to be
larger for the proton distribution than for the neutron one,
depending of course on the specific value of the ratio $Y_p/Y_n$.
As we will see later, this fact influences the values of the
contributions of protons and neutrons to the total entropy per
particle. The entropy, according to  relation (\ref{s-den-1}), is
an increasing function  of the diffuseness. Thus, the contribution
of each species on the total value of the entropy depends strongly
on the diffuseness of the corresponding Fermi distribution
function.

We can provide a second test, concerning the accuracy of the
present calculations, by comparing the entropy per baryon
calculated by applying Eqs.~(\ref{s-den-1}) and  (\ref{S-dif-f}).
Thus, in Fig.~15 we plot  $S$ of asymmetric nuclear matter with
$Y_p=0.2$ at $T=10,20,30$ MeV. The full lines give the entropy
calculated from Eq.~(\ref{s-den-1}), while the squares give
results obtained by differentiating $F(n,T)$ with respect to $T$
(Eq.~(\ref{S-dif-f})). It is obvious that there is again a  very
good agreement of the results, especially for low values of $T$
and high values of $n$.

In Fig.~16 we plot the contribution of the proton $S_p$, the
neutron $S_n$ and the total entropy per baryon $S$. It is obvious
that there is a strong effect of $T$ on the values of the
entropies mainly for low values of the density. The main part of
the contribution comes from neutrons, whereas the contribution of
protons is three times less. It is worthwhile to notice that, in
spite of $Y_p\sim (1/20-3/10)Y_n$, the approximate relation
$S_p\sim (1/4-3/7)S_n$ holds. This feature is understood by the
previous discussion that $f_p(n,T)$ is diffused more broadly than
$f_n(n,T)$, so the larger the diffuseness, the larger is the
entropy contribution (see also \cite{Takatsuka-94}).

In Fig.~17 we plot the contribution of the electronic $S_e$,  the
muonic $S_{\mu}$ and the total (leptonic) $S_l$ to the entropy per
baryon. The contribution to the entropy, of $S_e$ depends slightly
on the density, for fixed values of $T$. Our present results are
very close to those found by Onsi {\it et al.} \cite{Onsi-94},
where they employed the analytical approximate formula for the
electron entropy density $s_e$
\begin{equation}
s_e=\frac{1}{3}\frac{\mu_e^2}{(\hbar c)^3} T, \qquad \qquad
\mu_e=\hbar c(3\pi^2 Y_e n)^{1/3}. \label{Se-Onsi}
\end{equation}
According to the above formula, the contribution of electrons to
the entropy per baryon has the form
\begin{equation}
S_e=s_e/n \sim \left(\frac{Y_e^2}{n}\right)^{1/3}T.
\label{Se-Onsi-2}
\end{equation}
The quantity $\left(\frac{Y_e^2}{n}\right)^{1/3}$ is a  function
slightly dependent on the density $n$, so that for a fixed value
of $T$ the contribution $S_e$ is almost constant. The muonic
contribution to the entropy, for fixed $T$, increases slightly as
a function of the density.

In Fig.~18  we present the EOS of the $\beta$-stable hot nuclear
matter by taking into account and analyzing  the contribution to
the total pressure of each component. The main contribution to the
total pressure originates from the baryons, while the contribution
of the leptons is about a few percent compared to $P_b$. It is
worthwhile to notice that thermal effects are not important for
the calculation of $P_e$, but only for $P_{\mu}$, especially for
small values of $n$ ($n<0.4$ fm$^{-3}$). We found that thermal
effects produce a slightly stiffer equation of state with respect
to the case of cold nuclear matter. The above EOS can be applied
to the evaluation of the bulk properties of hot neutron stars
(mass and radius).

The study of hot nuclear matter in the absence of neutrino
trapping is the first step to study the properties of hot neutron
stars and supernova matter. Next, one can study the more realistic
case of neutrino-trapped matter in $\beta$-equilibrium. In this
case, the $\beta$-equilibrium conditions in  matter are altered
from the case in which neutrinos have left the system and thus the
composition of matter is affected. The proton fraction increases
dramatically and influences significantly the properties of
nuclear matter. Such a work is in progress.

\section{Summary}
The evaluation of the equation of state of hot nuclear matter is
an important problem in nuclear physics and astrophysics. EOS is
the basis ingredient for the study of the supernova explosion as
well as on determining  the properties of  hot neutron stars.  The
motivation of the present work is to apply a momentum-dependent
interaction model for the study of the hot nuclear matter EOS
under $\beta$-equilibrium in order to be able to study
simultaneously thermal effects, not only on the kinetic part of
the symmetry energy and symmetry free energy,  but also on the
interaction part of the above quantities as well. We calculate the
proton fraction, as well as the lepton fractions, by applying the
constraints for chemical equilibrium and charge neutrality. The
free energy, the internal energy and also the pressure are
calculated as  functions of baryon density and for various values
of temperature. We also concentrate on the evaluation of thermal
effects on the internal energy and baryon pressure. Special
attention is dedicated to the study of the contribution of the
components of $\beta$-stable nuclear matter on the entropy per
particle, a quantity of great interest in the study of structure
and collapse of supernova. We present and analyze the contribution
of each component. Finally, we present the EOS of $\beta$-stable
hot nuclear matter, by taking into account and analyzing  the
contributions to the total pressure of each component. The above
EOS can be applied to the evaluation of the gross properties of
hot neutron stars i.e. mass and radius, (work in progress).

\section*{Acknowledgments}
One of the authors (Ch.C.M) would like to thank Professor
Tatsuyauki Takatsuka for valuable comments and correspondence.

\section*{Appendix}
The energy density of baryons (Eq.~(\ref{E-D-1})), at $T=0$, is
given by
\begin{eqnarray}
\epsilon(n,I,T=0)&=&\frac{3}{10}E_F^0n_0u^{5/3}\left[(1+I)^{5/3}+(1-I)^{5/3}\right]+
\frac{1}{3}An_0\left[\frac{3}{2}-(\frac{1}{2}+x_0)I^2\right]u^2
\nonumber \\ &+&
\frac{\frac{2}{3}Bn_0\left[\frac{3}{2}-(\frac{1}{2}+x_3)I^2\right]u^{\sigma+1}}
{1+\frac{2}{3}B'\left[\frac{3}{2}-(\frac{1}{2}+x_3)I^2\right]u^{\sigma-1}}
 \\
&+&\frac{3}{2}n_0u\sum_{i=1,2}\left[C_i+\frac{C_i-8Z_i}{5}I\right]\left(\frac{\Lambda_i}{k_F^0}\right)^3
\left(\frac{\left((1+I)u\right)^{1/3}}{\frac{\Lambda_i}{k_F^0}}-
\tan^{-1} \frac{\left((1+
I)u\right)^{1/3}}{\frac{\Lambda_i}{k_F^0}}\right)\nonumber \\
&+&
\frac{3}{2}n_0u\sum_{i=1,2}\left[C_i-\frac{C_i-8Z_i}{5}I\right]\left(\frac{\Lambda_i}{k_F^0}\right)^3
\left(\frac{\left((1-I)u\right)^{1/3}}{\frac{\Lambda_i}{k_F^0}}-
\tan^{-1}
\frac{\left((1-I)u\right)^{1/3}}{\frac{\Lambda_i}{k_F^0}}\right)
\nonumber. \label{e-T0}
\end{eqnarray}
The pressure of the baryons, at $T=0$, defined as
\[P=n^2\frac{d(\epsilon/n)}{d n},\] is given by
\begin{eqnarray}
P(n,I,T=0)&=&\frac{1}{5}n_0E_F^0
u^{5/3}\left[(1+I)^{5/3}+(1-I)^{5/3}\right]+\frac{1}{3}n_0u^2A\left[\frac{3}{2}-\left(\frac{1}{2}+x_0\right)I^2\right]
\nonumber \\
\\
&+&\frac{2}{3}B\sigma n_0 u^{\sigma
+1}\frac{\left[\frac{3}{2}-(\frac{1}{2}+x_3)I^2\right]\left(1+\frac{2}{3\sigma}B'u^{\sigma
-1}\left[\frac{3}{2}-(\frac{1}{2}+x_3)I^2\right]
\right)}{\left(1+\frac{2}{3}B'\left[\frac{3}{2}-(\frac{1}{2}+x_3)I^2\right]u^{\sigma-1}\right)^2}
\nonumber \\
&+&
\frac{n_0u^2}{2}\sum_{i=1,2}\left[C_i+\frac{C_i-8Z_i}{5}I\right]
\left(\frac{\Lambda_i}{k_F^0}\right)^2
\frac{(1+I)^{1/3}}{u^{2/3}}\left(1-\frac{1}{1+\frac{(1+I)^{2/3}u^{2/3}}{\left(\frac{\Lambda_i}{k_F^0}\right)^2}}\right)
\nonumber \\
&+&\frac{n_0u^2}{2}\sum_{i=1,2}\left[C_i-\frac{C_i-8Z_i}{5}I\right]
\left(\frac{\Lambda_i}{k_F^0}\right)^2
\frac{(1-I)^{1/3}}{u^{2/3}}\left(1-\frac{1}{1+\frac{(1-I)^{2/3}u^{2/3}}{\left(\frac{\Lambda_i}{k_F^0}\right)^2}}\right).
\nonumber \label{P-all-T0}
\end{eqnarray}
In Eq.~(\ref{e-T0})  $E_F^0$ is the Fermi energy of symmetric
nuclear matter corresponding to equilibrium density $n_0$ and is
given by
\begin{equation}
E_F^0=\frac{\left(\hbar k_F^0\right)^2}{2m},\qquad k_F^0=\left(3
\pi^2\frac{n_0}{2}  \right)^{1/3}. \label{Ef-0}
\end{equation}
The Fermi momenta of protons and neutrons are
\[k_F^{p}=\left(3 \pi^2x n  \right)^{1/3}=\left( 3 \pi^2\frac{1-I}{2} n \right)^{1/3},\]
\[k_F^{n}=\left(3 \pi^2(1-x) n  \right)^{1/3}=\left(3 \pi^2\frac{1+I}{2} n  \right)^{1/3}.\]

%%%%%%%%%%%%%%%%%%%%%%%%%%%%%%%%%%%%
The chemical potentials of protons and neutrons, at $T=0$,  are
given by
\begin{eqnarray}
\mu_n&=&E+u\left(\frac{\partial E}{\partial
u}\right)_{Y_p}-Y_p\left(\frac{\partial E}{\partial Y_p}\right)_n,
\nonumber
\\ \mu_p&=&\mu_n+\left(\frac{\partial E}{\partial
Y_p}\right)_n, \\
\hat{\mu}&=&\mu_n-\mu_p=-\left(\frac{\partial E}{\partial
Y_p}\right)_n, \nonumber
 \label{mu-p}
\end{eqnarray}
where
\begin{eqnarray}
\left(\frac{\partial E}{\partial
u}\right)_{Y_p}&=&\frac{1}{5}E_F^{0}u^{-1/3}\left[(1+I)^{5/3}+(1-I)^{5/3}\right]
+
\frac{1}{3}A\left[\frac{3}{2}-(\frac{1}{2}+x_0)I^2\right]\nonumber\\
&+&
\frac{\frac{2}{3}Bu^{\sigma-1}\sigma\left[\frac{3}{2}-(\frac{1}{2}+x_3)I^2\right]
\left(1+\frac{2}{3\sigma}B'u^{\sigma-1}\left[\frac{3}{2}-(\frac{1}{2}+x_3)I^2\right]
\right) }
{\left(1+\frac{2}{3}B'u^{\sigma-1}\left[\frac{3}{2}-(\frac{1}{2}+x_3)I^2\right]
\right)^2} \nonumber \\
 &+&\frac{1}{2}\sum_{i=1,2}\left[C_i+\frac{C_i-8Z_i}{5}I\right]
\left(\frac{\Lambda_i}{k_F^0}\right)^2 \frac{(1+I)^{1/3}}{u^{2/3}}
\left(1-\frac{1}{1+\frac{(1+I)^{2/3}u^{2/3}}{\left(\frac{\Lambda_i}{k_F^0}\right)^2}}\right)
\nonumber \\
&+&\frac{1}{2}\sum_{i=1,2}\left[C_i-\frac{C_i-8Z_i}{5}I\right]
\left(\frac{\Lambda_i}{k_F^0}\right)^2 \frac{(1-I)^{1/3}}{u^{2/3}}
\left(1-\frac{1}{1+\frac{(1-I)^{2/3}u^{2/3}}{\left(\frac{\Lambda_i}{k_F^0}\right)^2}}\right),
%\nonumber
\end{eqnarray}
\begin{eqnarray}
\left(\frac{\partial E}{\partial
Y_p}\right)_n&=&\frac{1}{2}E_F^{0}u^{2/3}\left[(1+I)^{2/3}-(1-I)^{2/3}\right]-
\frac{1}{3}Au(\frac{1}{2}+x_0)I \nonumber\\
&-&
\frac{\frac{2}{3}Bu^{\sigma}(\frac{1}{2}+x_3)I}{\left(1+\frac{2}{3}B'u^{\sigma-1}\left[\frac{3}{2}-(\frac{1}{2}+x_3)I^2\right]
\right)^2} \nonumber\\
&+&\frac{3}{2}\sum_{i=1,2}\left[\frac{C_i-8Z_i}{5}I\right]\left(\frac{\Lambda_i}{k_F^0}\right)^3
\left(\frac{\left((1+I)u\right)^{1/3}}{\frac{\Lambda_i}{k_F^0}}-
\tan^{-1} \frac{\left((1+
I)u\right)^{1/3}}{\frac{\Lambda_i}{k_F^0}}\right)\nonumber \\
&-&
\frac{3}{2}\sum_{i=1,2}\left[\frac{C_i-8Z_i}{5}I\right]\left(\frac{\Lambda_i}{k_F^0}\right)^3
\left(\frac{\left((1-I)u\right)^{1/3}}{\frac{\Lambda_i}{k_F^0}}-
\tan^{-1}
\frac{\left((1-I)u\right)^{1/3}}{\frac{\Lambda_i}{k_F^0}}\right)
\nonumber \\
&+&\frac{1}{2}\sum_{i=1,2}\left[C_i+\frac{C_i-8Z_i}{5}I\right]
\left(\frac{\Lambda_i}{k_F^0}\right)^2 \frac{u^{1/3}}{(1+I)^{2/3}}
\left(1-\frac{1}{1+\frac{(1+I)^{2/3}u^{2/3}}{\left(\frac{\Lambda_i}{k_F^0}\right)^2}}\right)
\nonumber \\
&-&\frac{1}{2}\sum_{i=1,2}\left[C_i-\frac{C_i-8Z_i}{5}I\right]
\left(\frac{\Lambda_i}{k_F^0}\right)^2 \frac{u^{1/3}}{(1-I)^{2/3}}
\left(1-\frac{1}{1+\frac{(1-I)^{2/3}u^{2/3}}{\left(\frac{\Lambda_i}{k_F^0}\right)^2}}\right).
%\nonumber
\end{eqnarray}

%%%%%%%%%%%%%%%%%%%%%%%%%%%%%%%%%%%%%%
%%%%%%%%%%%%%%%%%%%%%%%%%%%%%%%%%%%%%%%
%%%%%%%%%%%%%%%%%%%%%%%%%%%%%%%%%%%%%%%%%%%%%%%%%%%%%
\newpage
%\FIGURE-1
\begin{figure}
\centering
\includegraphics[height=8.0cm,width=8cm]{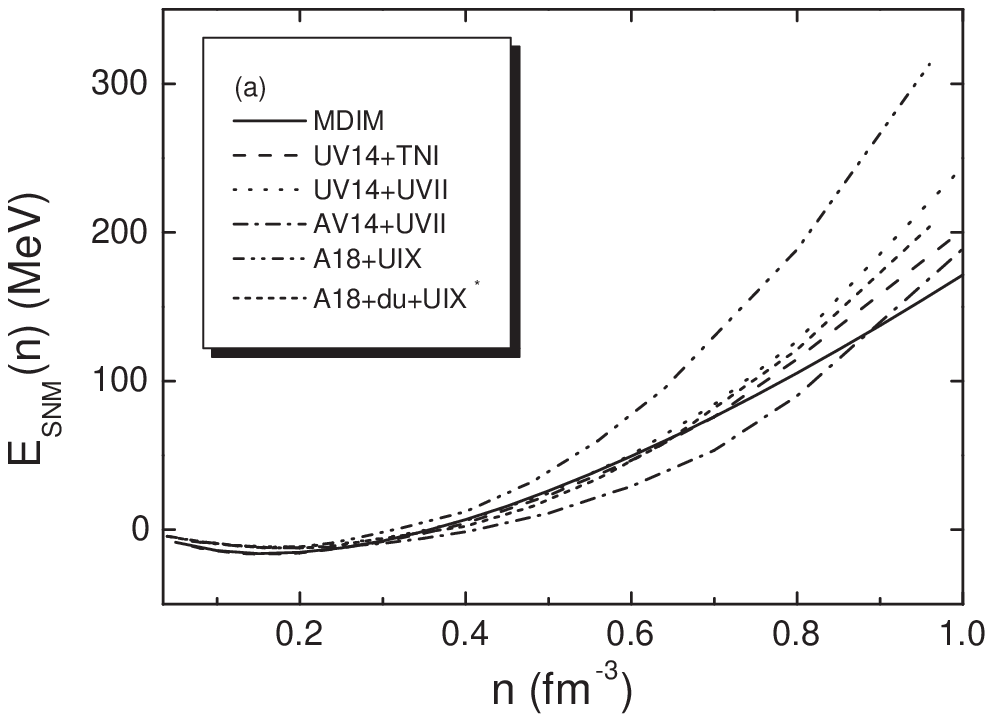}
 \includegraphics[height=8cm,width=8cm]{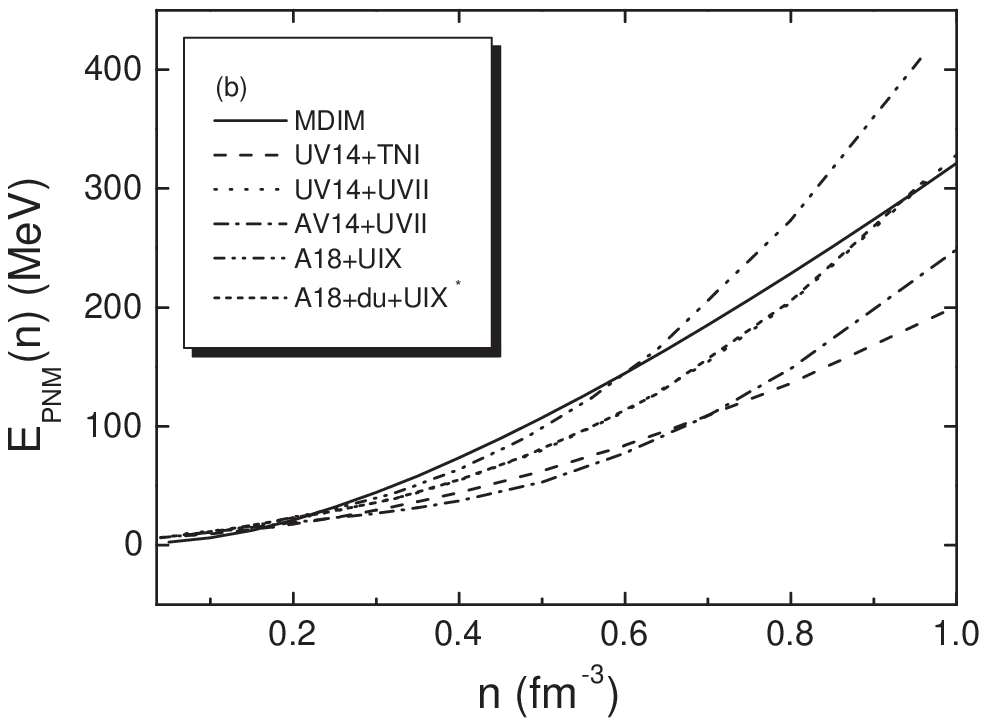}
\caption{ Energy per baryon of symmetric (a) and pure neutron
matter (b) of the present model (MDIM) in comparison with those
originated from realistic calculations. More details for the
models UV14+TNI, UV14+UVII and  AV14+UVII in
Ref.~\cite{Wiringa-88-b} and for the models A18+UIX and
A18+du+UIX$^*$ in Ref.~\cite{Pandharipande-98}. } \label{}
\end{figure}
%%%%%%%%%%%%%%%%%%%%%%%%%%%%%%%%%%%%%%%%%%%%%%%%%%%%%
%\newpage
%\FIGURE-2
\begin{figure}
\centering
\includegraphics[height=8.0cm,width=8cm]{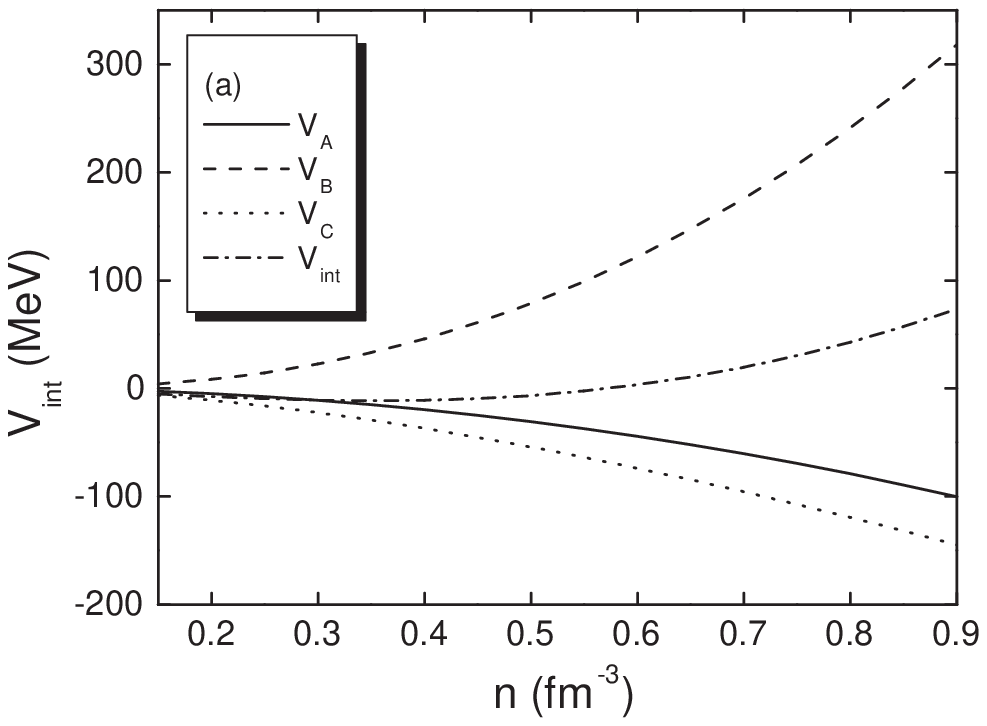}
 \includegraphics[height=8cm,width=8cm]{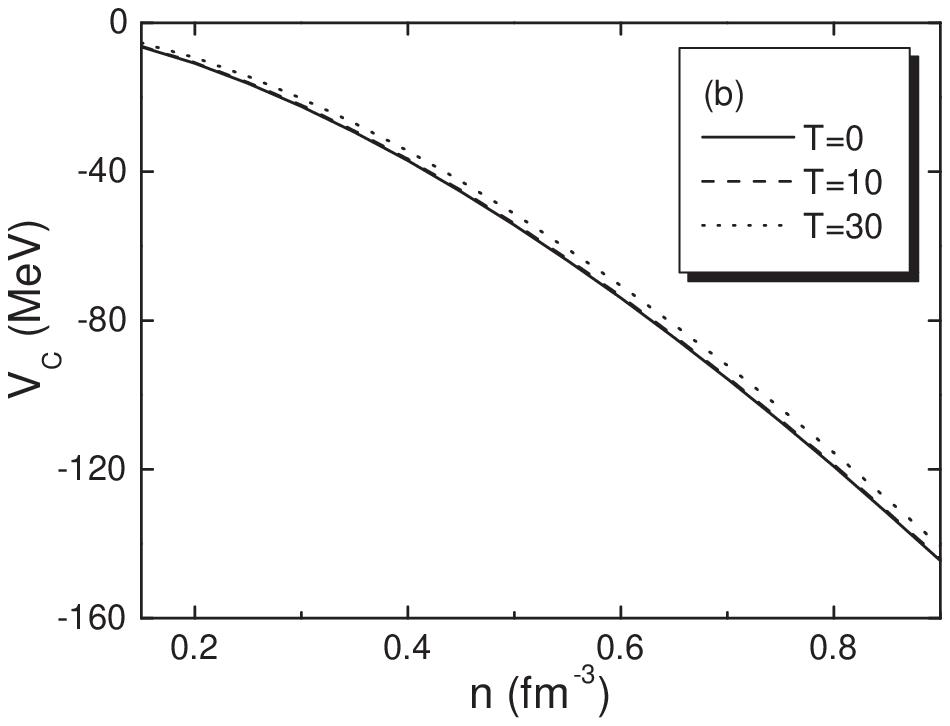}
\caption{a) The contribution of the various terms $V_A$, $V_B$,
$V_C$ and the total potential energy density $V_{int}$ as a
function of the baryon density  b) The momentum dependent term
$V_C$ as a function of the baryon density at temperature $T=0$,
$T=10$ and $T=30$ MeV. } \label{}
\end{figure}
%%%%%%%%%%%%%%%%%%%%%%%%%%%%%%%%%%%%%%%%%%%%%%%%%%%%%%%%%%%%%%%%%%%
%%%%%%%%%%%%%%%%%%%%%%%%%%%%%%%%%%%%%%%%%%%%%%%%%%%%%%%%%%%%%%%%%%%
%\newpage
%FIGURE-3
\begin{figure}
\centering
\includegraphics[height=7.0cm,width=5.7cm]{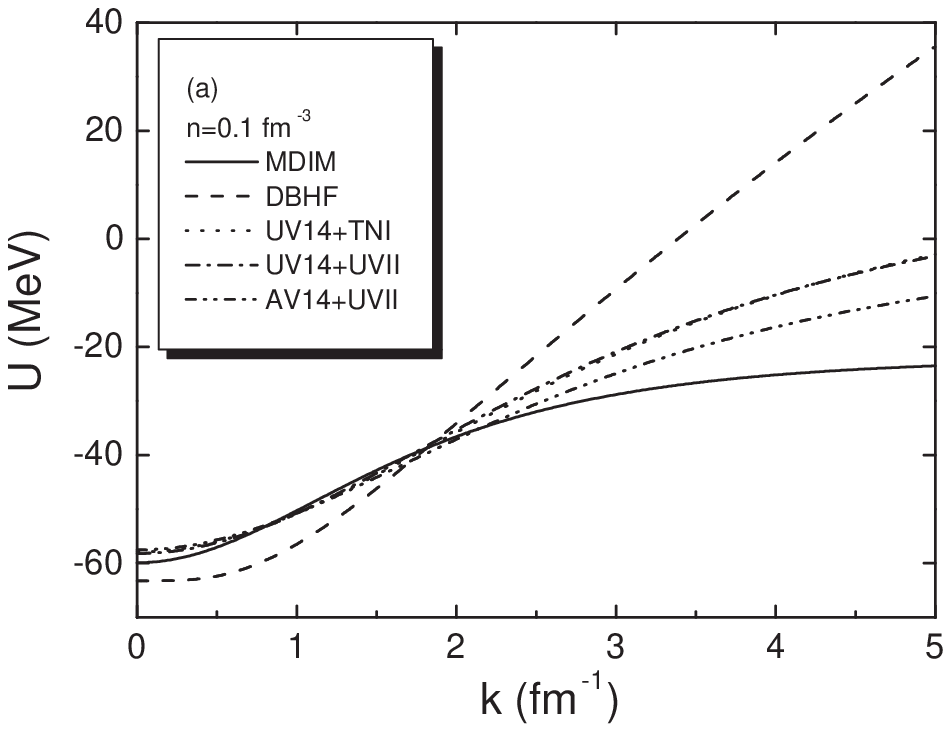}
 \includegraphics[height=7.0cm,width=5.7cm]{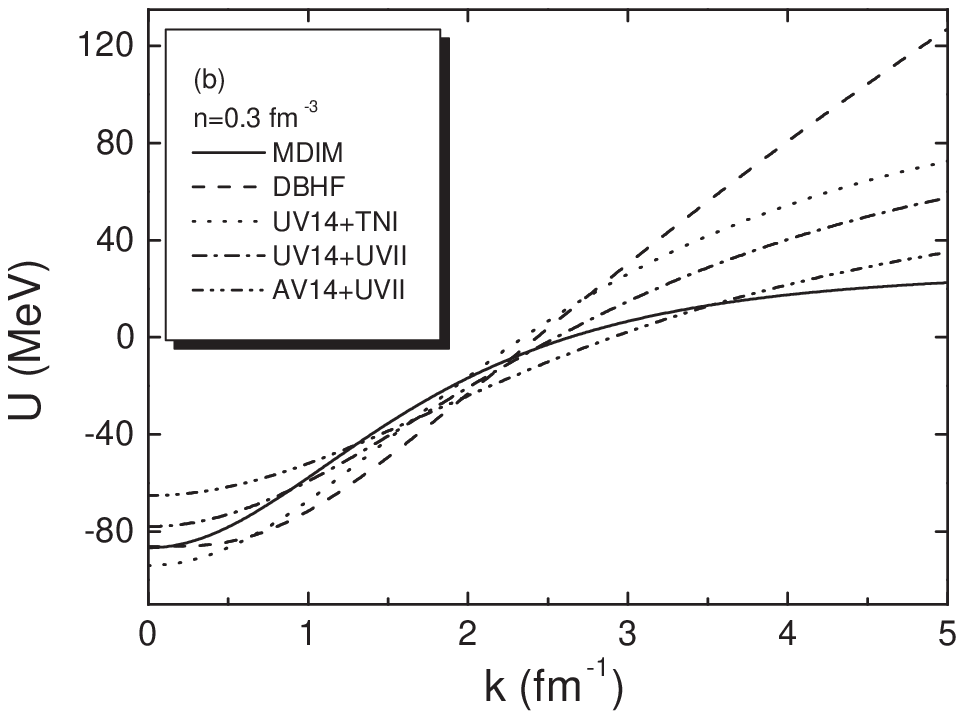}
 \includegraphics[height=7.0cm,width=5.7cm]{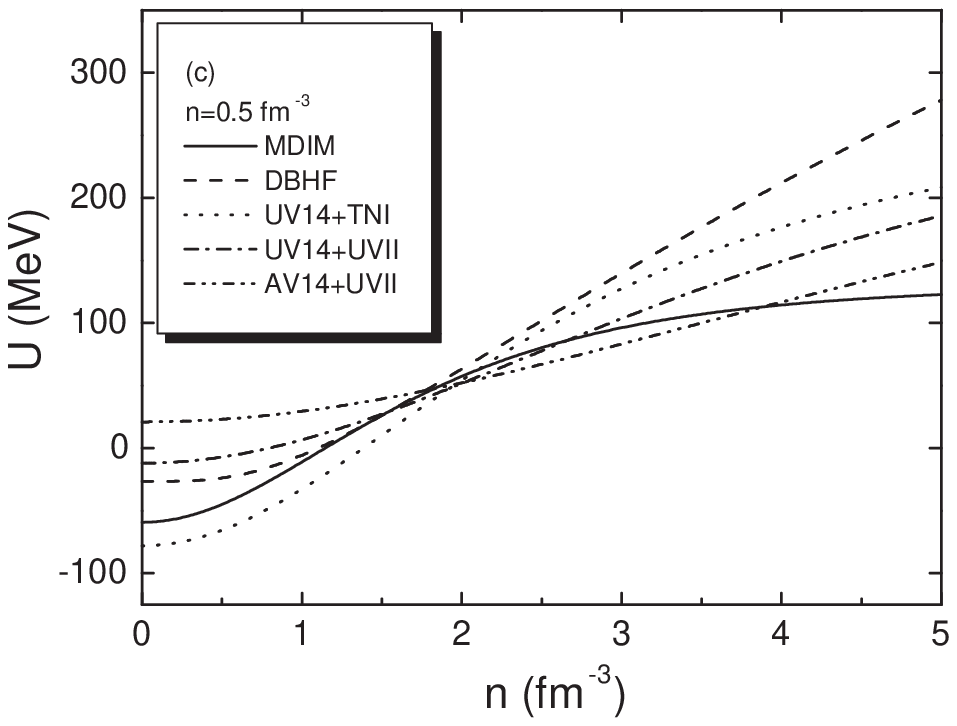}
\caption{A comparison of the single-particle potential of
symmetric nuclear matter from the present model (MDIM) with the
microscopic calculations of Wirigna~\cite{Wiringa-88-2} and Li
{\it et al}.~\cite{Li-93}, for densities $n=0.1$ fm$^{-3}$,
$n=0.3$ fm$^{-3}$  and $n=0.5$ fm$^{-3}$. } \label{}
\end{figure}

%%%%%%%%%%%%%%%%%%%%%%%%%%%%%%%%%%%%%%%%%%%%%%%%%%%%%%%%%%%%%%%%%%%%%%
%FIGURE-4
\begin{figure}
\centering
\includegraphics[height=8.0cm,width=8cm]{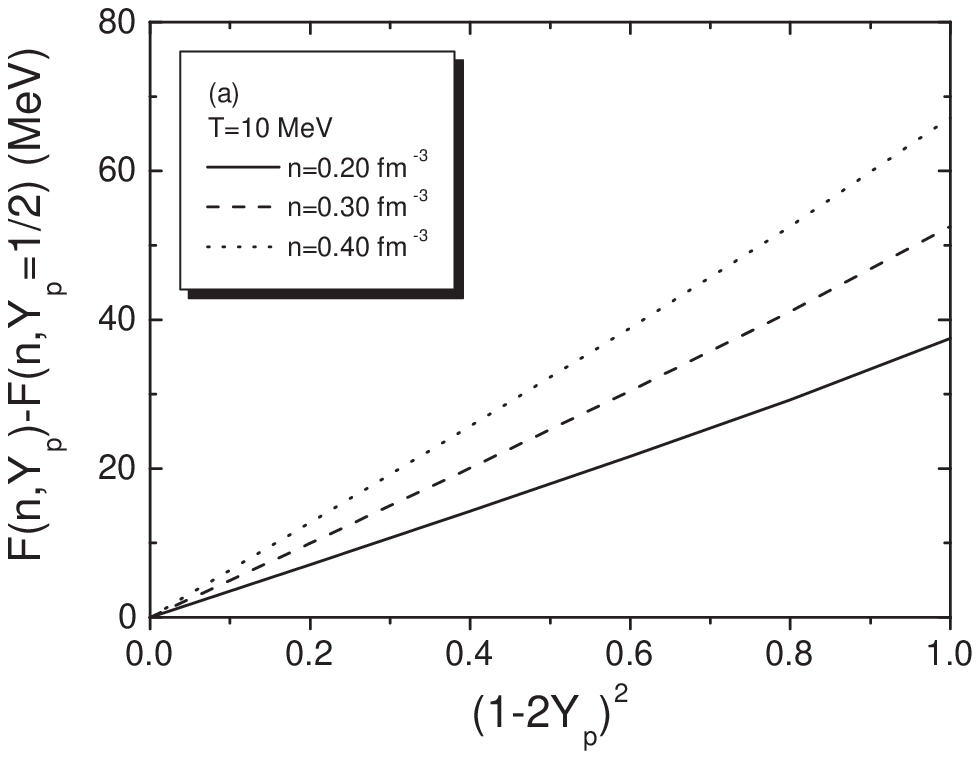}
 \includegraphics[height=8cm,width=8cm]{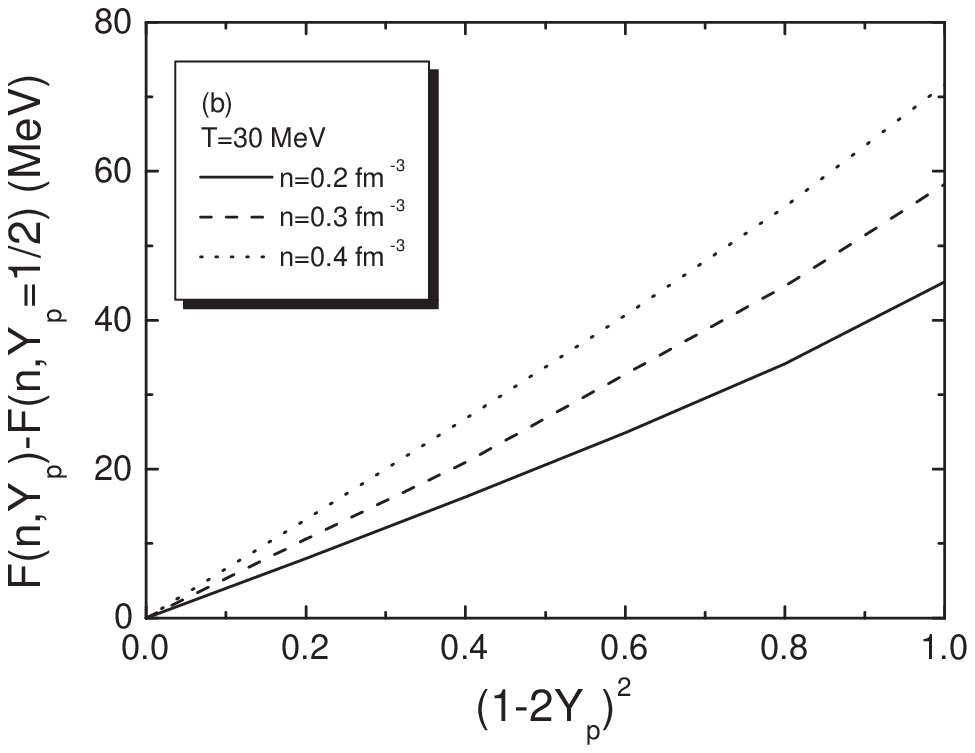}
\caption{The difference $F(n,T,Y_p)-F(n,T,Y_p=1/2)$ as a function
of $(1-2Y_p)^2$ at temperatures a) $T=10$ and b) $T=30$ MeV, for
three baryon densities. } \label{}
\end{figure}
%%%%%%%%%%%%%%%%%%%%%%%%%%%%%%%%%%%%%%%%%%%%%%%%%%%%%%%%%%%%%%%%%%%%%%%%
%FIGURE-5
\begin{figure}
\centering
\includegraphics[height=9.0cm,width=9cm]{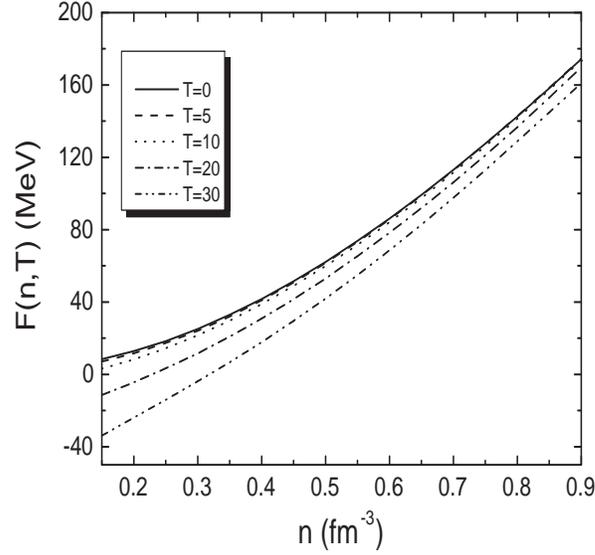}
\caption{The Helmholtz free energy $F(n,T,I)$ of $\beta$-stable
matter versus the baryon density $n$, for various values of $T$
(in MeV). } \label{}
\end{figure}
%%%%%%%%%%%%%%%%%%%%%%%%%%%%%%%%%%%%%%%%%%%%%%%%%%%%%%%%%%%%%%%%%%%%%%%%
%FIGURE-6
\begin{figure}
\centering
\includegraphics[height=8.0cm,width=8cm]{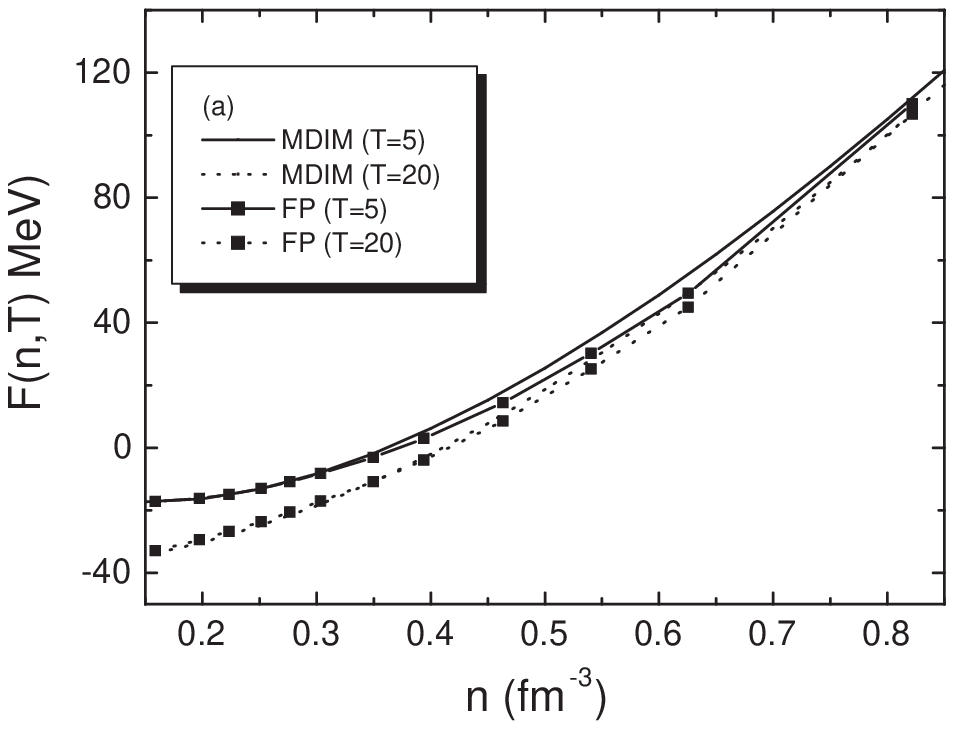}
\includegraphics[height=8.0cm,width=8cm]{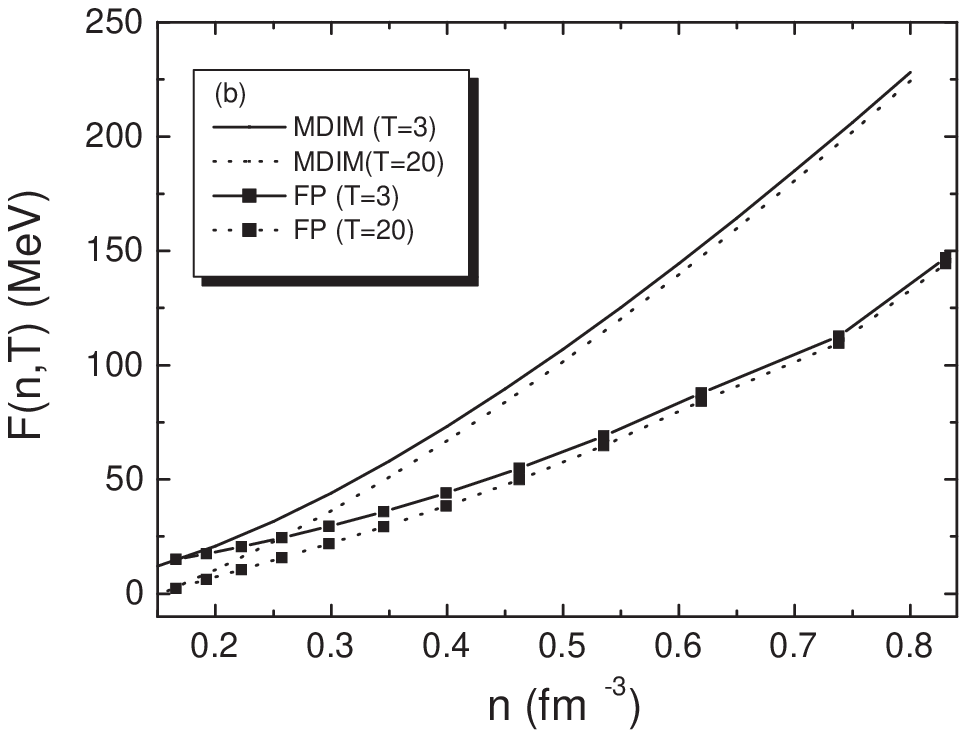}
\caption{(a) The free energy of  symmetric nuclear matter (for
$T=5$ MeV and $T=20$ MeV) and (b) for pure neutron matter (for
$T=3$ MeV and $T=20$ MeV) of the proposed model (MDIM) in
comparison with the free energy calculated by Friedman and
Pandharipande model (FP) \cite{Friedman-81}. } \label{}
\end{figure}
%%%%%%%%%%%%%%%%%%%%%%%%%%%%%%%%%%%%%%%%%%%%%%%%%%%%%%%%%%%%%%%%%%%%%%%%
%FIGURE-7
\begin{figure}
\centering
\includegraphics[height=9.0cm,width=9cm]{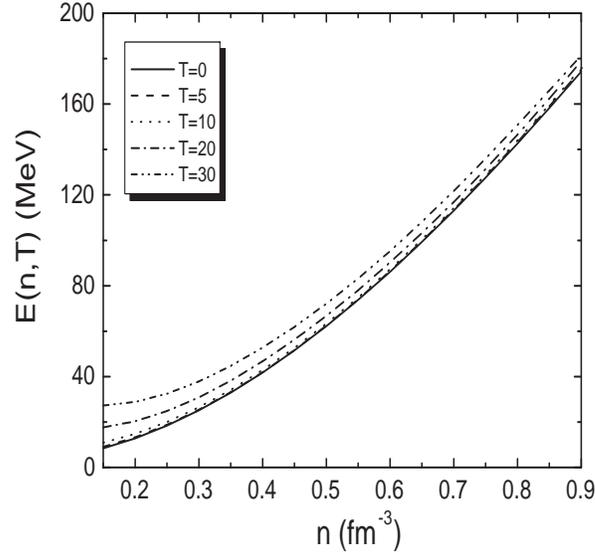}
\caption{The internal energy $E(n,T)$ of $\beta$-stable matter as
a function of the baryon density $n$ for various values of $T$. }
\label{}
\end{figure}
%%%%%%%%%%%%%%%%%%%%%%%%%%%%%%%%%%%%%%%%%%%%%%%%%%%%%%%%%%%%%%%%%%%%%%%%
%%%%%%%%%%%%%%%%%%%%%%%%%%%%%%%%%%%%%%%%%%%%%%%%%%%%%%%%%%%%%%%%%%%%%%%%
%FIGURE-8
\begin{figure}
\centering
\includegraphics[height=8.0cm,width=8cm]{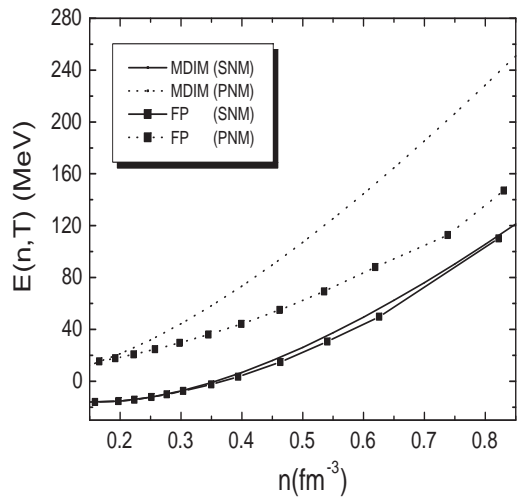}
\caption{The internal energy (for $T=0$ MeV) for  symmetric
nuclear matter (SNM) and pure neutron matter (PNM) calculated with
the MDIM in comparison with the FP model. } \label{}
\end{figure}
%%%%%%%%%%%%%%%%%%%%%%%%%%%%%%%%%%%%%%%%%%%%%%%%%%%%%%%%%%%%%%%%%%%%%%%%
%%%%%%%%%%%%%%%%%%%%%%%%%%%%%%%%%%%%%%%%%%%%%%%%%%%%%%%%%%%%%%%%%%%%%%%%
%FIGURE-9
\begin{figure}
\centering
\includegraphics[height=8.0cm,width=8cm]{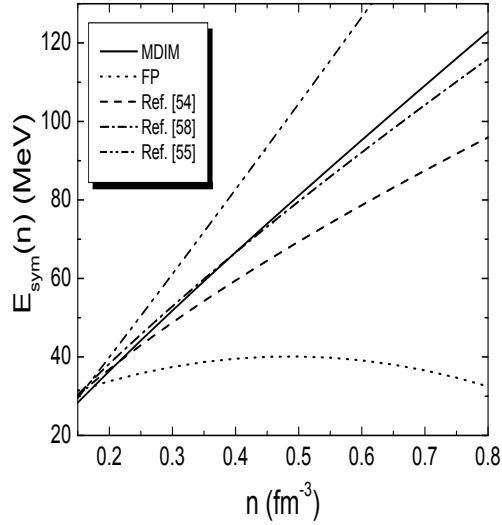}
\vspace{-3cm} \caption{The nuclear symmetry energy calculated with
the MDIM in comparison with the FP model as well as the results of
Refs.~\cite{Shetty-07},~\cite{Sammarruca-08} and \cite{Chen-05}. }
\label{}
\end{figure}
%%%%%%%%%%%%%%%%%%%%%%%%%%%%%%%%%%%%%%%%%%%%%%%%%%%%%%%%%%%%%%%%%%%%%%%%
%%%%%%%%%%%%%%%%%%%%%%%%%%%%%%%%%%%%%%%%%%%%%%%%%%%%%%%%%%%%%%%%%%%%%%%%
%FIGURE-10
\begin{figure}
\centering
\includegraphics[height=9.0cm,width=9cm]{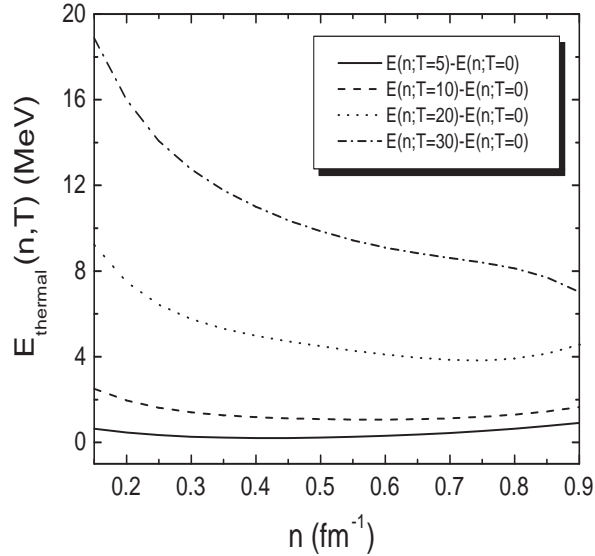}
\caption{The thermal energy $E_{thermal}(n,T)=E(n,T)-E(n,T=0)$ of
$\beta$-stable matter versus the baryon density $n$, for various
values of $T$. } \label{}
\end{figure}
%%%%%%%%%%%%%%%%%%%%%%%%%%%%%%%%%%%%%%%%%%%%%%%%%%%%%%%%%%%%%%%%%%%%%%%%
%FIGURE-11
\begin{figure}
\centering
\includegraphics[height=8.0cm,width=8cm]{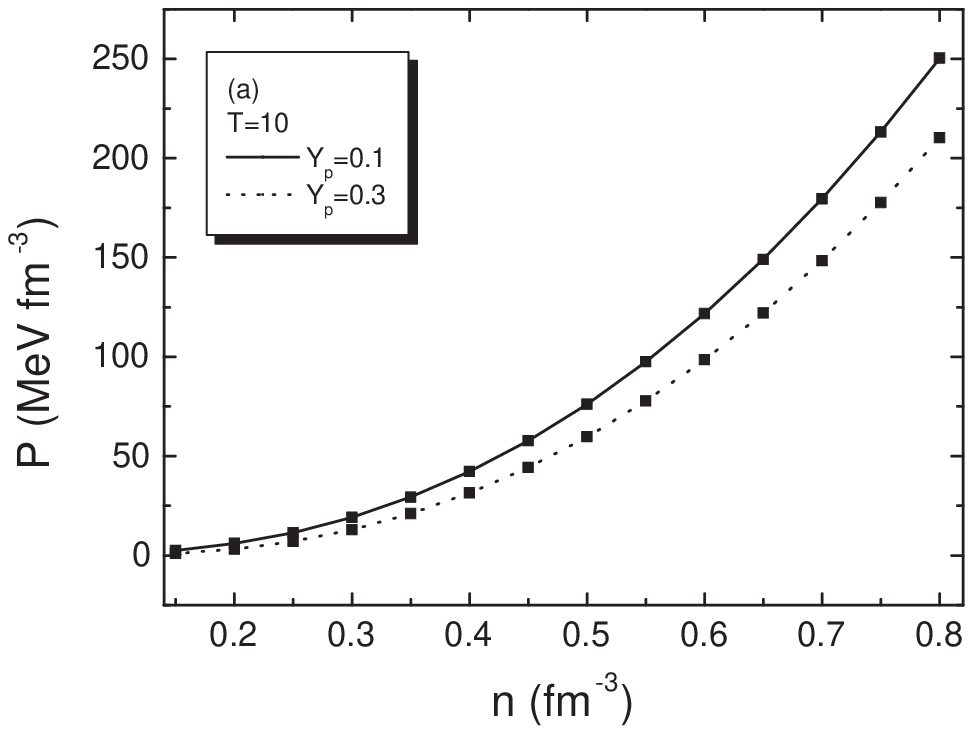}
\includegraphics[height=8.0cm,width=8cm]{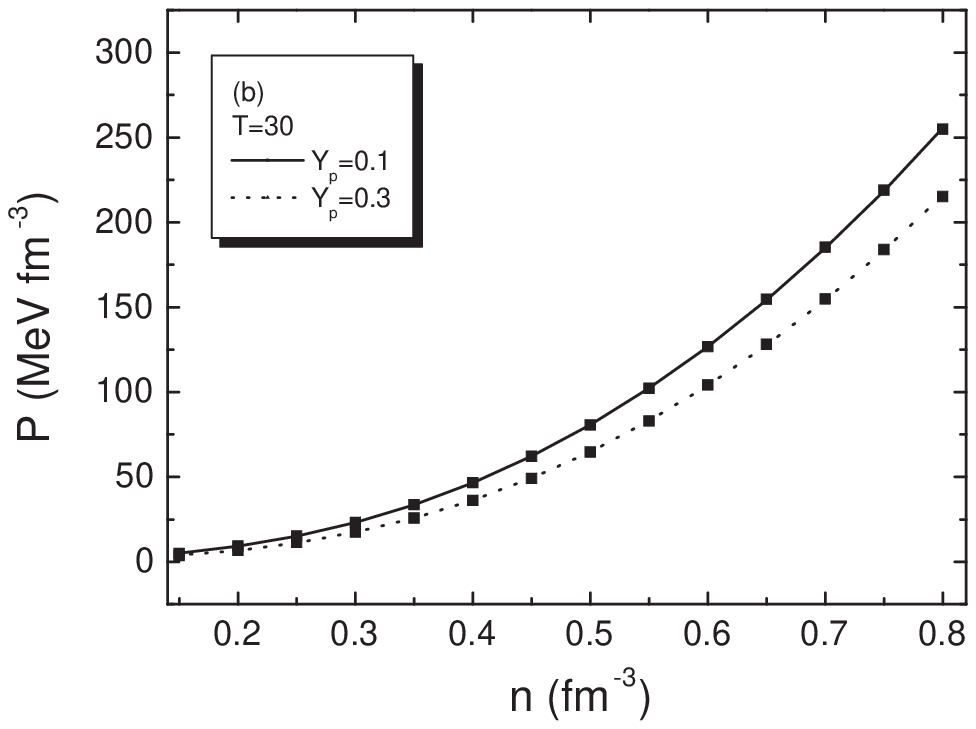}
\caption{The pressure $P$ of asymmetric nuclear matter for
$Y_p=0.1$ and $0.3$ at a) $T=10$ and b) $30$ MeV. The full lines
give the results calculated from Eq.~(\ref{P-1}), while the
squares represent results obtained by differentiating $F(n,T)$
(Eq.~(\ref{P-m-F})). } \label{}
\end{figure}
%%%%%%%%%%%%%%%%%%%%%%%%%%%%%%%%%%%%%%%%%%%%%%%%%%%%%%%%%%%%%%%%%%%%%%%%
%FIGURE-12
\begin{figure}
\centering
\includegraphics[height=9.0cm,width=9cm]{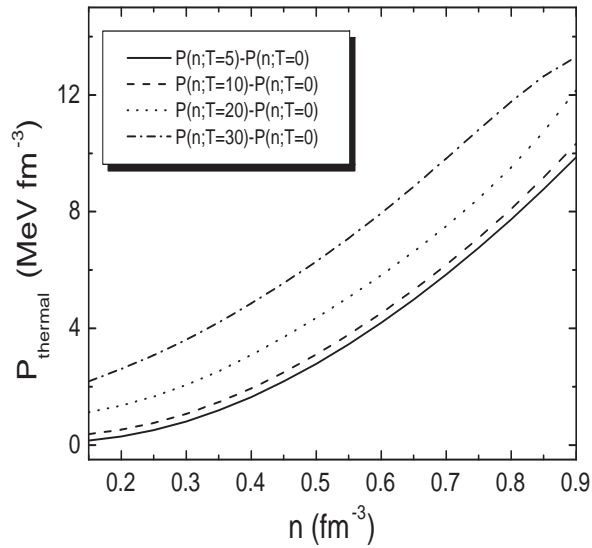}
\caption{The thermal pressure $P_{thermal}(n,T)=P(n,T)-P(n,T=0)$
of $\beta$-stable matter versus the baryon density $n$, for
various values of $T$. } \label{}
\end{figure}
%%%%%%%%%%%%%%%%%%%%%%%%%%%%%%%%%%%%%%%%%%%%%%%%%%%%%%%%%%%%%%%%%%%%%%%%
%FIGURE-13
\begin{figure}
\centering
\includegraphics[height=8.0cm,width=8.cm]{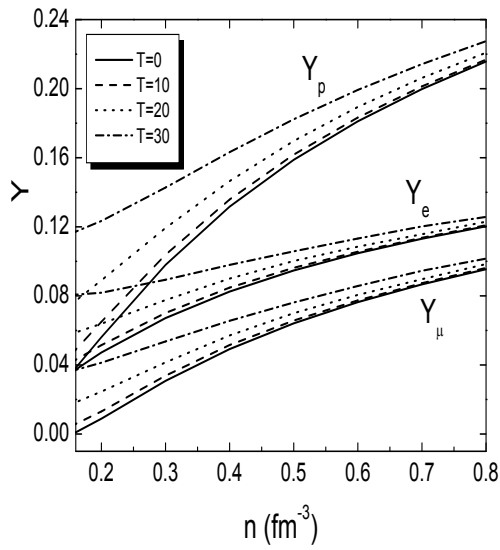}\
\vspace{-3.2cm} \caption{The fractions of protons $Y_p$ electrons
$Y_e$ and muons $Y_{\mu}$ of $\beta$-stable matter as functions of
the baryon density $n$, for various values of  $T$. } \label{}
\end{figure}
%%%%%%%%%%%%%%%%%%%%%%%%%%%%%%%%%%%%%%%%%%%%%%%%%%%%%%%%%%%%%%%%%%%%%%%
%%%%%%%%%%%%%%%%%%%%%%
%FIGURE-14
\begin{figure}
\centering
\includegraphics[height=6.0cm,width=5.5cm]{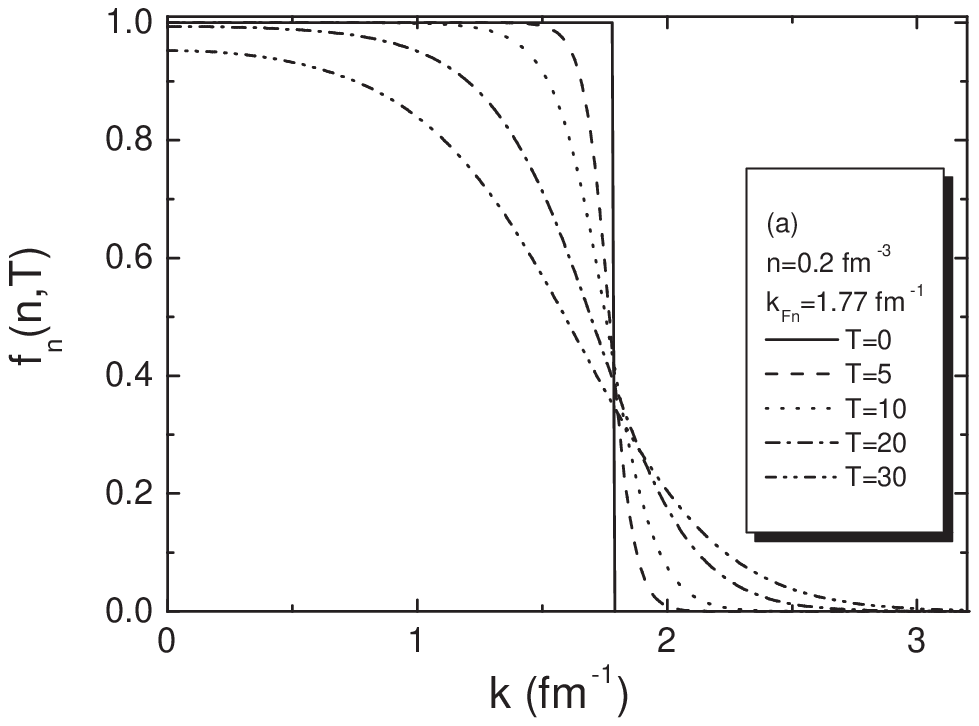}\
\includegraphics[height=6.0cm,width=5.5cm]{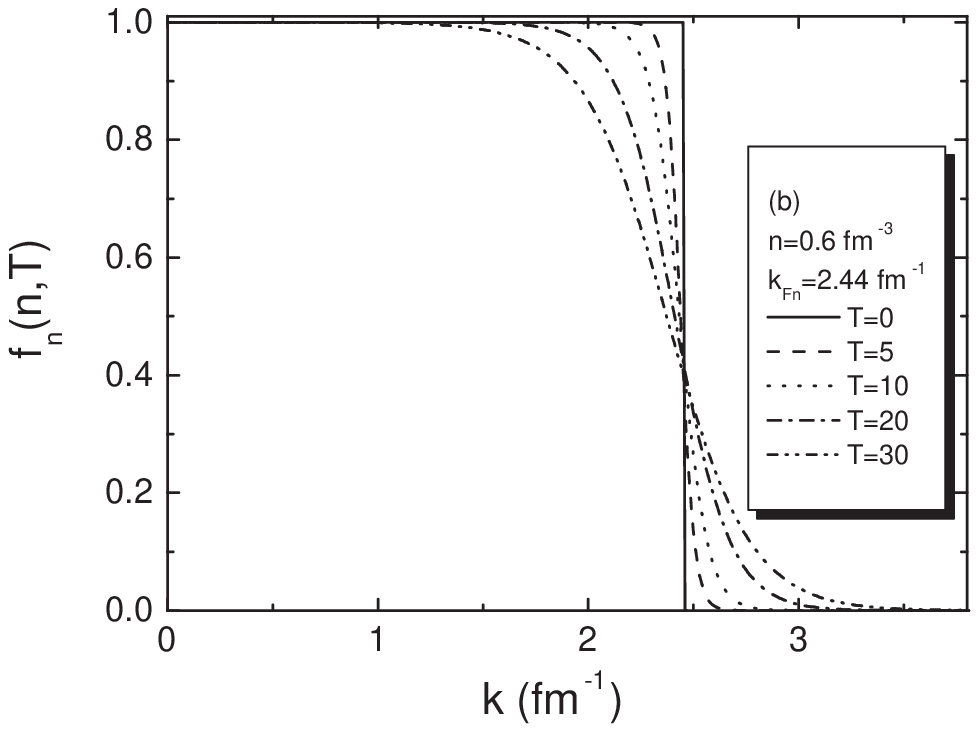}\\
\includegraphics[height=6.0cm,width=5.5cm]{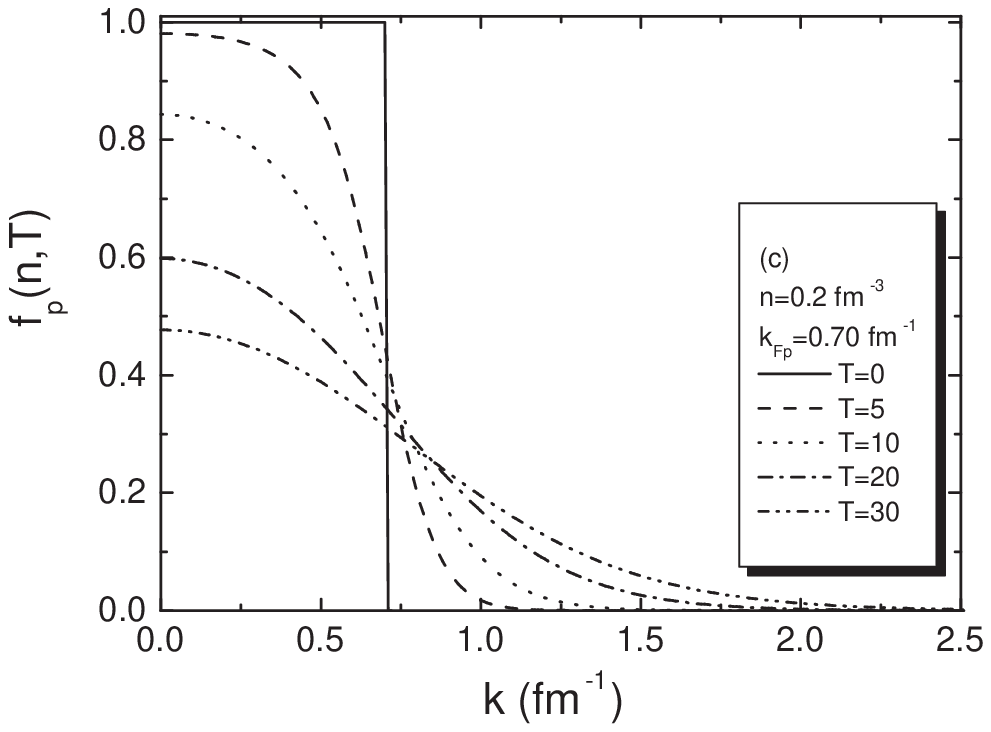}\
\includegraphics[height=6.0cm,width=5.5cm]{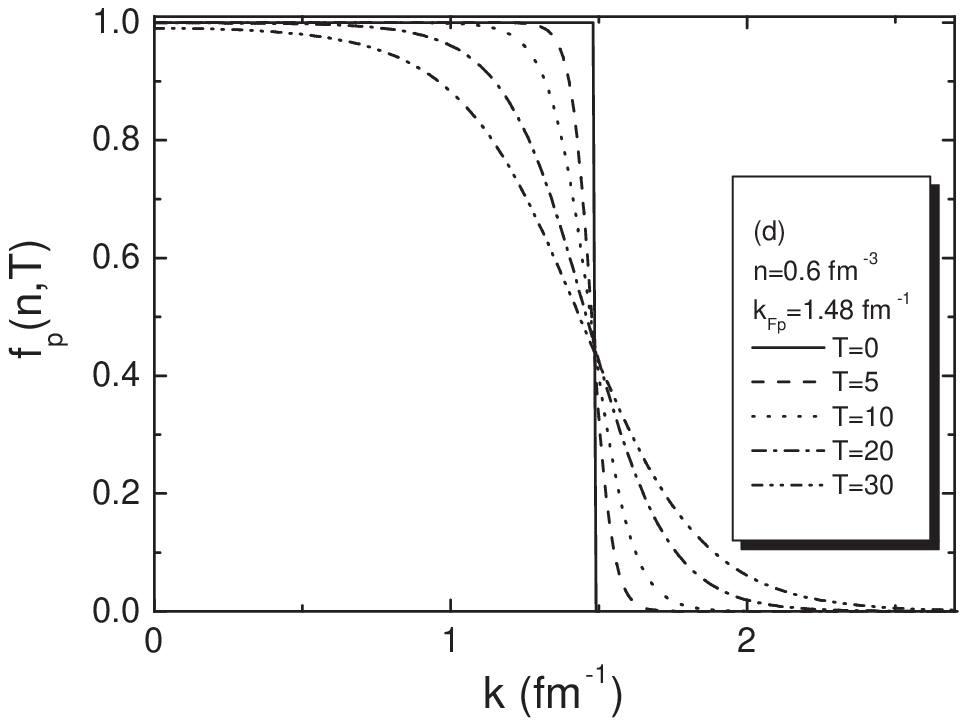}\
\caption{The Fermi-Dirac distribution function $f_{\tau}(n,T)$ for
protons and neutrons ($\tau=p,n$ respectively), for $n=0.2$
fm$^{-3}$, $n=0.4$ fm$^{-3}$ and $n=0.6$ fm$^{-3}$ and various
values of $T$. } \label{}
\end{figure}
%%%%%%%%%%%%%%%%%%%%%%%%%%%%%%%%%%%%%%%%%%%%%%%%%%%%%%%%%%%%%%%%%%%%%%%
%FIGURE-15
\begin{figure}
\centering
\includegraphics[height=9.0cm,width=9cm]{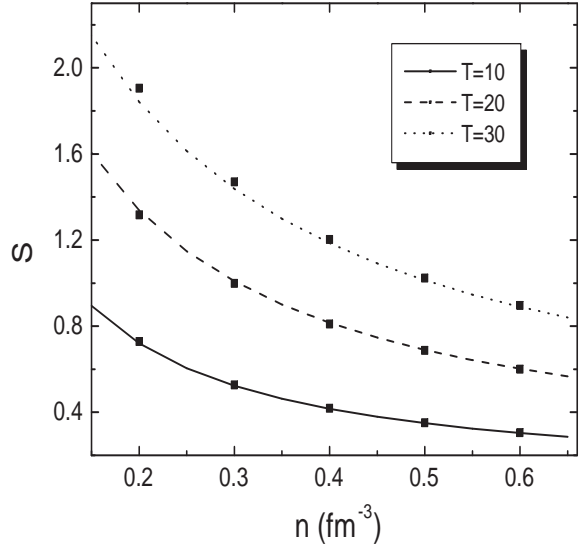}
\caption{The entropy per particle $S$ of asymmetric nuclear matter
with  $Y_p=0.2$ at $T=10,20,30$ MeV. The full lines give the
entropy calculated from Eq.~(\ref{s-den-1}), while the squares
give results obtained by differentiating $F(n,T)$
(Eq.~(\ref{S-dif-f})). } \label{}
\end{figure}
%%%%%%%%%%%%%%%%%%%%%%%%%%%%%%%%%%%%%%%%%%%%%%%%%%%%%%%%%%%%%%%%%%%%%%
%FIGURE-16
\begin{figure}
\centering
\includegraphics[height=8.0cm,width=8.0cm]{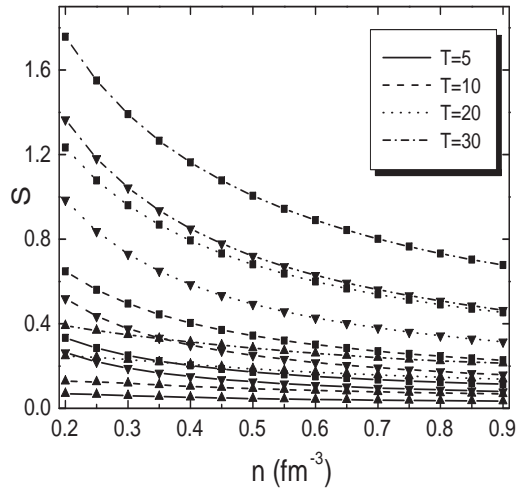}\
\caption{Contributions to the total entropy per particle of
protons ($S_p$) (up triangles)  neutrons ($S_n$) (upside down
triangles)  and the total entropy ($S_b$) (squares).} \label{}
\end{figure}
%%%%%%%%%%%%%%%%%%%%%%%%%%%%%%%%%%%%%%%%%%%%%%%%%%%%%%%%%%%%%%%%%%%%%%
%%%%%%%%%%%%%%%%%%%%%%%%%%%%%%%%%%%%%%%%%%%%%%%%%%%%%%%%%%%%%%%%%%%%%%%%
%FIGURE-17
\begin{figure}
\centering
\includegraphics[height=8.0cm,width=8.0cm]{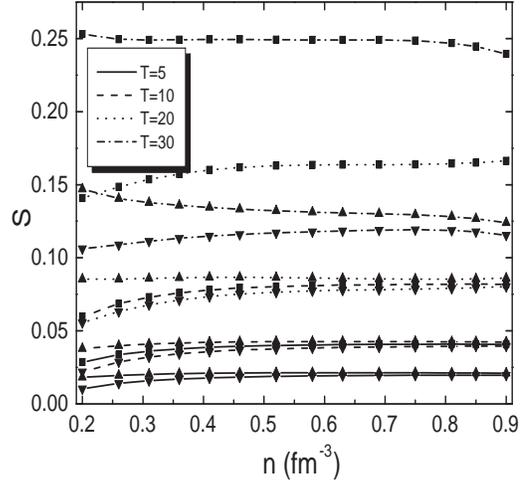}\
%\vspace{-3.2cm}
 \caption{Contributions to the total entropy per
particle of  electrons ($S_e$) (up triangles)  muons ($S_{\mu}$)
(upside down triangles) and the total $S_l$ (squares).} \label{}
\end{figure}
%%%%%%%%%%%%%%%%%%%%%%%%%%%%%%%%%%%%%%%%%%%%%%%%%%%%%%%%%%%%%%%%%%%%%%%
%%%%%%%%%%%%%%%%%%%%%%%%%%%%%%%%%%%%%%%%%%%%%%%%%%%%%%%%%%%%%%%%%%%%%%%%
%FIGURE-18
\begin{figure}
\centering
\includegraphics[height=8.0cm,width=8.0cm]{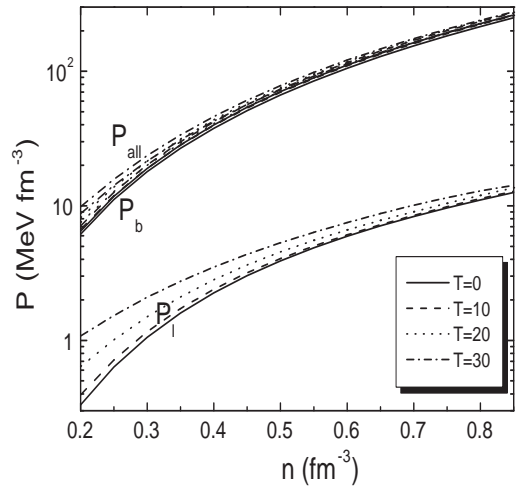}\
\caption{The pressures of baryons $P_b$ leptons $P_l$
(electrons+muons) and the total pressure $P$ versus the baryon
density, $n$ for various values of $T$.} \label{}
\end{figure}

%%%%%%%%%%%%%%%%%%%%%%%%%%%%%%%%%%%%%%%%%%%%%%%%%%%%%%%%%%%%%%%%%%%%%%%

\end{document}